# T.I.P.O.
# (Tesla Interferometric Planetary Observer)

## Space Mission Proposal

Author: Adrian Sabin Popescu
Astronomical Institute of the Romanian Academy
e-mail: sabinp@aira.astro.ro

# 1. Introduction

In the last years the Space Science community was confronted to a continuous increasing interest in Martian missions, extra-solar planet search and multi-satellite missions.

The here presented T.I.P.O. mission is a proposal for a research program dedicated to study, by space borne interferometric methods, the radio emissions generated in the atmospheres and magnetospheres of planets, both solar and extra-solar.

In the first sections of this material we will give an insight of the present knowledge of radio emissions from planetary atmospheres and magnetospheres in the Extremely Low Frequency (ELF) and Very Low Frequency (VLF) bands, intended to justify the relevance and necessity of a mission studying these phenomena. The next sections will be dedicated to the description of the T.I.P.O. scientific objective, how it is intended to achieve these objective, and to the ELFSAT satellites instrumentation.

## 2. ELF Creation and Propagation in Planetary Resonant Cavities

Earth can be regarded as a nearly conducting sphere, wrapped in a thin dielectric atmosphere that extends up to the ionosphere, whose conductivity is also substantial. Therefore, the surface and the ionosphere of Earth form a cavity where electromagnetic waves propagate. When a cavity is excited with broadband electromagnetic sources, a resonant state can develop provided the average equatorial circumference is approximately equal to an integral number of wavelengths of the electromagnetic waves. The propagation of ELF waves within such cavities was first studied by Schumann (1952) and the associated resonance phenomena were subsequently observed by Balser and Wagner (1960). Resonances in the Earth cavity are closely related with lightning activity and contain information about the global electric circuit of the cavity. Their characterization contributes not only to our understanding of wave propagation but also to the study of the **atmospheric composition** and of the **physical and chemical processes** that take place in the atmosphere. In certain conditions, the same theory can be used to study other planetary cavities, for example the ones of rocky planets, icy moons and gaseous giants, where the wavelengths are commensurate with planetary sizes and atmospheric compositions. In a first stage we can study through their VLF and ELF emissions our solar system planets and satellites (Venus, Mars, Jupiter and its moons Io and Europa, Saturn and its largest moon Titan, Uranus and Neptune) and, for a sufficient emission power and sensitive antennas to indirectly observe other planetary systems, of interest being to put into evidence by other than optical means the existence of extrasolar rocky planets, their size and, hopefully, to model their atmospheric composition.

The approximate theory of ELF propagation in the Earth ionosphere transmission line described by Booker (1980) is applied to a simplified worldwide model of the D and E regions, and the Earth's magnetic field (Behroozi-Toosi and Booker, 1983). At 1000 Hz by day, reflection is primarily from the gradient on the underside of the D region. At 300 Hz by day, reflection is primarily from the D region at low latitudes, but is from the E region at high latitudes. Below 100 Hz by day, reflection is primary from the gradient on the underside of the E region at all latitudes. By night, reflection from the gradient on the topside of the E region is important. There is then a resonant frequency (~ 300 Hz) at which the optical thickness of the E region for the whistler mode is half a wavelength. At the Schumann resonant frequency in the Earth-ionosphere cavity (7.53 Hz) the nocturnal E region is almost completely transparent for the whistler mode and is semi-transparent for the Alfvén mode. Reflection then takes place from the F region. ELF propagation in the Earth-ionosphere transmission line by night is quite dependent on the magnitude of the drop in ionization density between the E and F regions. Nocturnal propagation at ELF therefore depends significantly on an ionospheric feature whose magnitude and variability.

Fluctuations in the horizontal stratification in the ionosphere occur in both space and time, and they cause the direction of propagation in the transmitted ELF wave to be slightly spread. The effect is important because the wavelengths of radio waves in the reflecting stratum at ELF are comparable with wavelengths involved in acoustic gravity waves in the atmosphere. What is needed is, not the behavior of the transmitted wave in the unique direction calculated on the basis of strict horizontal stratification, but an average behavior for directions within a cone whose angle is controlled by the fluctuations in the angle of tilt of the stratification.

### 3. ELF and VLF Creation and Propagation in Planetary Magnetospheres

### 3.1. Outer Magnetospheres

The Sun's ionized corona expands rapidly, nearly isotropic, into the surrounding heliosphere carrying the solar magnetic field with it. This expanding magnetic plasma interacts with the planets creating planetary magnetospheres, regions of enhanced magnetic field strength surrounding the planet where magnetic stresses play a dominant role in controlling the flow of plasma. The nature of the obstacle to the solar wind flow is important in determining the nature of a planetary magnetosphere. If the planet has its own strong magnetic field due to an interior magnetic dynamo or a remnant field of a magnetized crust, then the magnetosphere is termed intrinsic. The size of an intrinsic magnetosphere is determined by the location of the point where the dynamic pressure of the solar wind flow is equal to the magnetic pressure of the intrinsic magnetic field. For Mercury this occurs just above the surface of the planet, whereas for Jupiter it occurs close to 100 times further from the center of the planet than the cloud tops. Because the solar wind flows supersonically, that is the flow is much faster than the speed of a compressional wave that could deflect the plasma flow, then a bow shock forms that slows, heats and deflects the flow before it reaches the magnetic obstacle. This shock stands off in front of the obstacle at a distance sufficiently far that the compressed solar wind plasma can flow around the obstacle.

Induced magnetospheres occur when the region of strong magnetic field arises not from magnetism within the planetary obstacle but from the interaction of the solar wind with the obstacle. Induced magnetospheres in turn can be divided into two classes, one in which the planetary body is adding mass to the solar wind flow in the form of ions newly created from a neutral atmosphere. This occurs most spectacularly at comets and less so at Venus and Mars and also at the moons Io and Titan where the magnetosphere is created by a flowing planetary wind. The other class of induced magnetospheres arises from classical electromagnetic induction in a highly electrically conducting medium. This conductor can be the electrically conducting interior of a planet or moon such as an iron core or a salt-water ocean, such as in Europa, or it can be the ionosphere of a planet above the surface, such as at Venus and Mars. In these cases the magnetic field external to the planetary body is excluded from the interior of the conductor for a period of time dependent on the size and electrical conductivity of the obstacle. This time can be long compared to the time scale of directional changes in the exterior magnetic field. Currents flow in the conductor to oppose the field change and the field outside obstacle increases, creating a magnetic barrier that in turn deflects the flowing plasma. As in the case of intrinsic magnetospheres, either class of induced magnetospheres can lead to the formation of a bow shock, standing in front of the obstacle when the flow velocity exceeds the velocity of the compressional wave that is required to deflect the flow around the obstacle.

Herein we examine the physics of the outer magnetospheres. We have visited the magnetospheres of Jupiter and Saturn many times but those of Uranus and Neptune only once. These magnetospheres are all intrinsic.

Above we discussed briefly the size of intrinsic planetary magnetospheres as determined by a balance of pressures, between the dynamic pressure of the flowing solar wind and the magnetic pressure of the intrinsic magnetic field. Figure 1 shows an intrinsic magnetosphere as deduced from observations of the terrestrial magnetosphere. The magnetosphere forms a cavity in the flowing solar wind and throughout much of the cavity the pressure is dominated by the magnetic field. The pressure that determines the size of the cavity is normal to the surface of the cavity. Outside the cavity is plasma whose pressure decreases from outside to inside across the cavity boundary. This pressure gradient exerts an inward force. It is balanced by a magnetic pressure force pushing outward as the magnetic field increases from outside to inside across the cavity boundary, or **magnetopause**. The relationship between this normal component of the pressure and the directed dynamic pressure in the upstream solar wind is complex and is only understood semi-empirically (Petrinec and Russell, 1997). Similarly, the magnetic field depends on a multitude of currents, not just those interior to the planet. Thus the point of pressure equilibrium or force balance can move. Since the magnetic flux crossing the surface of the planet is fixed on the time scale of most magnetospheric events, any variations in the magnetopause location under conditions of constant external pressure must involve a change in shape of the magnetosphere. Rapidly rotating magnetospheres with strong internal sources of plasma add an additional complexity that will be discussed below.

The magnetosphere illustrated in Figure 1 is stretched in the antisolar direction. This **magnetotail**, as it is called, **is a region in which energy can be stored**, analogous to the energy that can be stored in an electrical circuit by an inductor. However, to understand the behavior of the magnetosphere one has to understand the transport of mass and magnetic flux that occur on time scales much slower than the speed of light. Thus one can often catch a magnetosphere in the act of changing, or in a metastable state, ready to change.

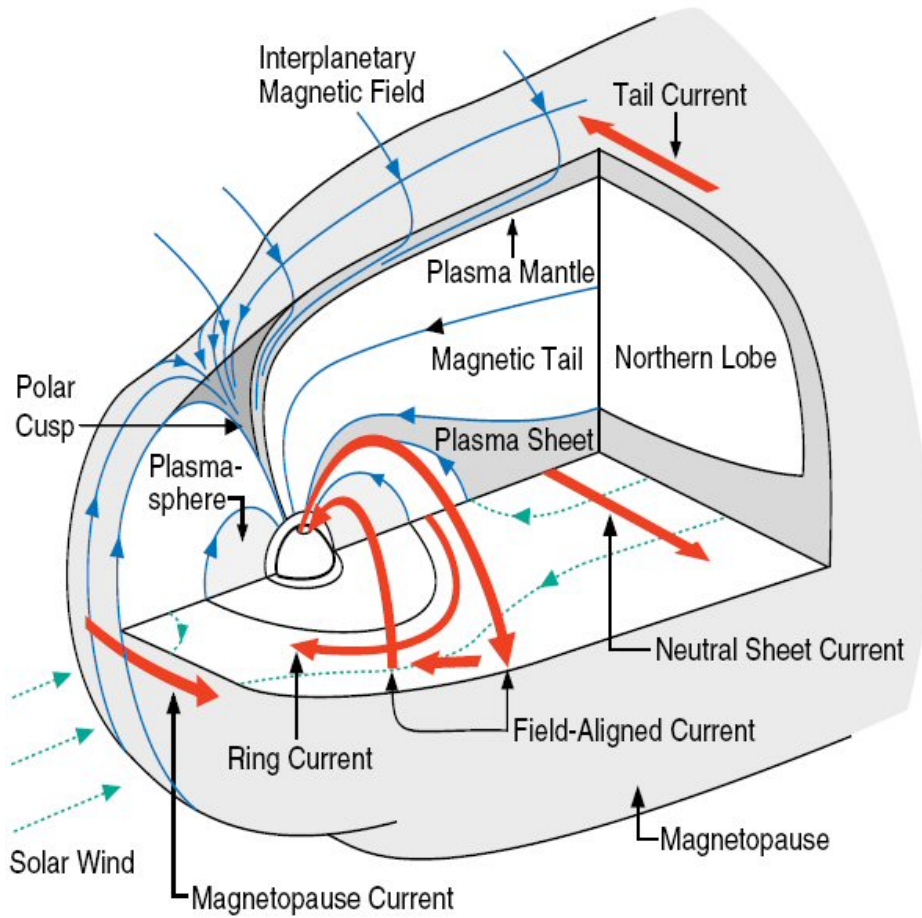

Figure 1: Cut away model of the terrestrial magnetosphere showing the plasma regions, magnetic field lines, electric currents and flows (Russell, 2004).

One can add energy to this system by simply compressing it, just as one can store energy in the compressed spring. The work done, or energy stored, is the force times the distance moved normal to the surface, integrated over the magnetosphere. One can also add energy to the system by "eroding" the magnetic field on the dayside and carrying it into the tail. This requires a tangential stress such as friction at the surface of the magnetosphere. This tangential stress can be a **viscous interaction associated with waves on the boundary in the presence of dissipation**, with particles being scattered from the flowing, shocked solar wind plasma into the magnetosphere, or by magnetic coupling of the magnetospheric and solar wind magnetic field. This coupling is illustrated in Figure 1 above the magnetosphere and behind the depression in the shape of the magnetopause called the polar cusp. Here the magnetic field lines are bent in the direction as to slow the solar wind flow. As the solar wind slows in this region, mechanical energy is removed from the flow and is stored in the tail lobes where the **magnetic field energy increases**. The transport of energy into the tail occurs through an electromagnetic Poynting flux (Russell, 2004).

Magnetospheres are vast in scale but contain very little total mass. The stress applied to the magnetosphere by the solar wind must be ultimately taken up by the planet. One of the ways to transmit this stress to the ionosphere and thence through collisions to the upper atmosphere and the planet, is through currents parallel to the magnetic field as illustrated in Figure 1. These currents close on pressure gradients in the magnetosphere. Another is through the gradient in the Chapman-Ferraro currents acting on the planet's dipole (Siscoe, 1966). This later process is most important on planets as Mercury where closure currents at the field of field lines may be inhibited.

Obviously it is important to understand the effective viscosity of the solar wind flow past the magnetosphere. This viscosity depends on many factors. Waves set up on the magnetopause by the non-steady interaction of the solar wind could act to drag on the magnetospheric plasma if the magnetosphere dissipated the waves set up on the magnetospheric side of the boundary. Similarly the conditions in the plasma exterior to the magnetopause, the magnetosheath, also affect the viscosity, especially if it occurs through the coupling of magnetic field across the boundary through the process that is called **magnetic reconnection**.

The energetic ions that have been created or reflected from the shock or leaking from the region behind the shock stream back into the solar wind leading to additional unsteadiness in the interaction on that side (generally the dawn side) of the magnetosphere. Behind the shock the plasma is slowed, compressed, heated and deflected, only to later expand as it moves behind the magnetosphere. We know how much slowing, compression, heating and deflection occurs via the Rankine-Hugoniot equations, a combination of fluid equations of motion and Maxwell's electromagnetic equations as applied to thin planar boundaries. These equations tell us that the strength of the bow shock is controlled by the speed of the fast mode compressional wave relative to the solar wind velocity. This ratio, called the Mach number, is greater than one when the solar wind velocity normal to the shock front exceeds the fast mode speed upstream at the shock. The greater is this ratio the hotter the plasma downstream becomes until the thermal pressure reaches the upstream dynamic pressure. Since the magnetopause is a region of pressure balance along the normal to the boundary, this increase in temperature with Mach number decreases the density (and the magnetic field) just exterior to the Magnetopause (Le and Russell, 1994). This weakening of the potential magnetic stress at the magnetopause appears also to weaken the coupling of the solar wind to the Earth's magnetosphere (Scurry and Russell, 1991). The variation of the Mach number and the beta value (the ratio of magnetic to thermal pressure) of the solar wind with the heliocentric distance is expected to affect processes in the solar wind but not so much reconnection at the outer planets because the beta value in the magnetosheath is dominated by the large Mach number of the preshock solar wind flow. The expected strength of the outer planet bow shocks is confirmed by their large overshoots in magnetic field strength, much exceeding overshoots seen in the strongest terrestrial bow shocks (Russell et al., 1982). Thus we expect that magnetic reconnection with the solar wind may play a much lesser role at the outer planets than at Mercury and the Earth.

### 3.2. Reconnection

The process known as reconnection (Dungey, 1961) leads to the connection of the terrestrial and solar wind magnetic fields. This has long been postulated as the ultimate cause of the geomagnetic storm, substorms and the circulation of plasma in the terrestrial magnetosphere. It occurs on the dayside of the terrestrial magnetosphere when the solar wind magnetic field is southward, opposite than of the terrestrial magnetic field at the subsolar point. It accelerates the plasma in the direction of the magnetic stress as expected from theory (Paschmann et al., 1979) and the process may be quasi-stationary (Sonnerup et al., 1981) or time-varying (Russell and Elphic, 1978). Spatially limited time-varying reconnection results in a phenomenon that has been termed a flux transfer event. In a flux transfer event a bipolar magnetic field signature appears in the direction along the magnetopause normal and a strengthened field tangential to the boundary as a flux rope slides along the magnetopause. Such flux transfer event are observed frequently at Earth and Mercury and perhaps less often (and weaker) at Jupiter (Russell, 2004).

Very few magnetopause data are available at the magnetopause crossings of the outer planets, Saturn, Uranus and Neptune, but the data that are available do suggest that reconnection occurs (Huddleston et al., 1997).

The presence of reconnection at the terrestrial magnetosphere is critical to the energization of magnetospheric processes. It is not obvious, however, that it should be as important in energizing the magnetospheres of the outer planets of the solar system. As mentioned above, high solar wind Mach numbers mitigate against reconnection. The Mach number in the outer solar system is about twice the value in the inner solar system and hence we expect that magnetopause reconnection would be weaker in the outer solar system (Russell, 2004).

A second argument for diminished efficiency of reconnection is the size of the magnetospheric standoff distance measured in terms of the ion inertial length. For the plasma conditions present at the planets this is basically the ion gyroradius. The success of this parameter in ordering the nature of physical processing in magnetospheres of different size (Omidi et al., 2003) is recognition of the importance of the ratio of the radius of curvature of the obstacle relative to the ion scale in controlling these processes. This value is large for all planets, thus justifying the frequent use of a fluid approximation, such as a magnetohydrodynamic (MHD) simulation, to treat large-scale processes at all the planets. However, this ratio does vary by an order of magnitude from Mercury to Jupiter. Since reconnection is a kinetic and not a fluid phenomenon, all else being equal we would expect reconnection to be most important at Mercury, Neptune and Uranus and least important at Jupiter. Simply put, the size of the region on the magnetopause in which kinetic effects take place (the neutral point) is much smaller at Jupiter relative to the dimension of the system than at the other planets, as a result of the large radius of curvature of the jovian magnetosphere.

For Jupiter and possibly for Saturn there is a third reason why we would not expect magnetopause reconnection to be important. There is much more effective engine for driving plasma circulation in these two magnetospheres and for energizing the magnetosphere. It is the same process, **centrifugal driven flow**, that is responsible for stretching the equatorial dimension for the jovian magnetosphere.

Ionospheres are coupled to planetary atmospheres and thence to the rotating planet and tend to rotate with the planet. The ionosphere in turn couple to the magnetosphere and cause the charged particles there to attempt to rotate with the ionosphere. At distances inside the synchronous orbit where particles would orbit the planet with the period less than that of the rotation of the planet, the inward force of gravity is too great for a corotating particle to remain in orbit so that an additional force (provided generally by the magnetic field) is required to support the particle. At distances outside the synchronous orbit the inward force of gravity is too weak for a corotating particle to remain in a corotating trajectory so an additional inward force is needed, again generally provided by the magnetic field.

Visualizing the behavior of particles in a rotating system is complicated if one does not move into the frame of reference of the rotating system. In general we do not consider centrifugal force when treating the Earth's magnetosphere even through the plasmasphere corotates with the Earth and is dense, relative to the outer magnetosphere. The reason we do not is because for the Earth the gravitational force and centrifugal force balance at 6.6 $R_E$. Here the plasma is not very dense in general and even is the plasma from 6.6 $R_E$ to the magnetopause were accelerated to corotational velocities the magnetic force would easily balance the centrifugal force. At slowly rotating Mercury, the synchronous distance at which a body would remain over the same location is 96 Mercury radii, far outside the magnetosphere.

Jupiter is at the other extreme. Synchronous orbit is at 2.3 jovian radii. There is mass loading body, the moon Io, at 5.9 jovian radii that adds a ton per second of plasma to the magnetosphere. The centrifugal force of corotating plasma at Io (and beyond) exceeds by far the gravitational force so that only the magnetic forces are available to confine the plasma. If the equatorial magnetic field cannot contain the plasma, the field will become more and more distorted. If the field of the field lines cannot be frozen into the ionosphere they will "slip". In either case the plasma will circulate in response to the mass loading process. At a rate of a ton per second it does not take long (of the order of months) to build up the plasma density at Io to the value observed and a value that forces the plasma to convect outward (which results in also a **radial instability** having consequences on wave-particle interactions that will be analyzed in a following section).

Synchronous orbit is similarly close to the planet at Saturn, Uranus and Neptune. However, Uranus and Neptune do not have obvious strong plasma sources within the magnetospheres. Saturn has a strong icy ring system extending well beyond synchronous orbit so that it can provide mass that could promote centrifugally driven circulation. Moreover, at great distances, ~ $R_S$, there is Titan, the moon with the greatest atmosphere of any in the solar system. However, here the magnetic field is very weak and the interaction between the mass loaded plasma and the magnetosphere may be much different than at Io.

Figure 2 illustrates how the magnetospheric flux tubes couple to the ionosphere and ultimately to the planetary ionosphere (giving also a way of **channeling the Schumann planetary cavity waves into the magnetosphere**). At high altitudes a set of magnetic field lines are pushed by the plasma, so that they are sheared with respect to the surrounding flux tubes. When the magnetic field lines are sheared, a field aligned current arises. This current closes in the resistive ionosphere and causes a Lorentz or **J**×**B** force that drags on the ionosphere. At the top of the flux tube in the magnetosphere the currents also close across the field lines flowing (radially in the case of Jupiter) on pressure gradient surfaces orthogonal to the pressure gradient force.

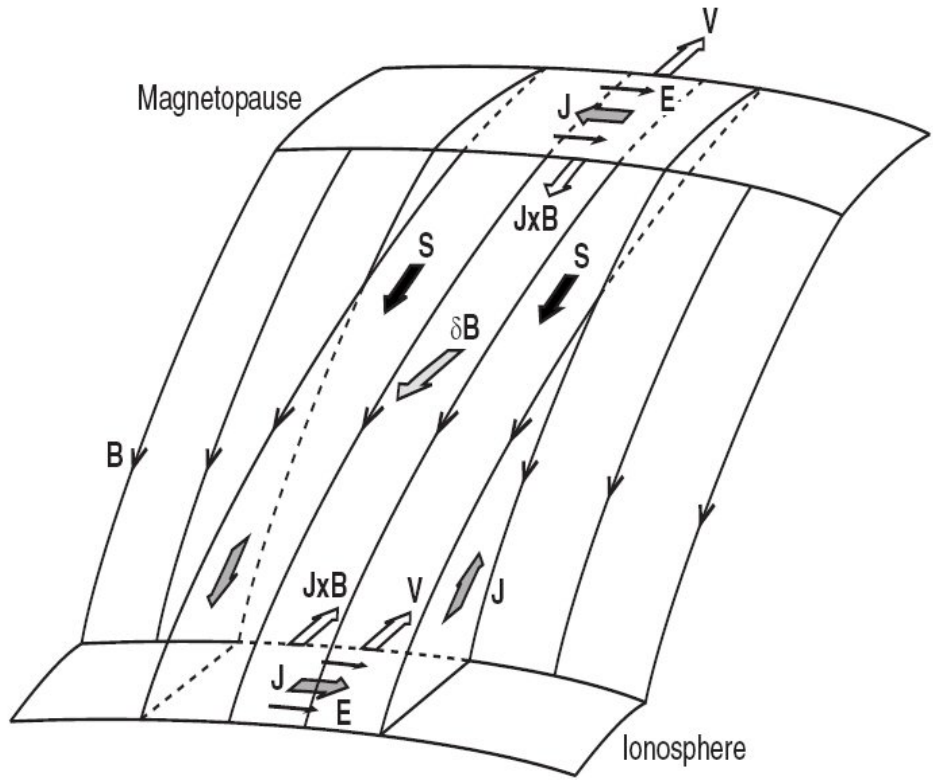

Figure 2: The transmission of stress between the ionosphere and magnetosphere. Stresses in either the ionosphere or magnetosphere can be transmitted between regions by shearing a flux tube with respect to its neighbors. Currents close in the ionosphere across magnetic field lines and couple the tube to the ionosphere plasma. Currents close in the magnetosphere via pressure gradient drift currents. Currents join the two regions along the magnetic field (Strangeway et al., 2000).

At Io, where the moon orbits at 17 km/s and the torus plasma rotates at 74 km/s, the upper atmosphere of Io becomes ionized and is accelerated by the electric field associated with the corotating plasma. On a kinetic level the newly added ions form a ring in velocity space that has free energy that can be released as **ion cyclotron waves** oscillating at the gyrofrequency of the newly created ion. The ion cyclotron wave transfers some energy from perpendicular to along the field and helps to isotropize the plasma. The mass loading region around Io was observed by Galileo spacecraft, observations that included also the measurement of the ion cyclotron wave amplitude (Huddleston, 1999).

### 3.3. Circulation of Plasma

It is clear from the disk-like shape of the jovian magnetosphere that the equatorial regions have been mass loaded and it is clear the in situ measurements at Io that the mass loading is occurring there as ions are picked up from the upper atmosphere by the corotating torus. We even understand how this plasma is forced to corotate by the coupling to the ionosphere but we have not yet examined how mass loading supplants the terrestrial dayside reconnection process as the driver of convection. Such a plasma circulation model was proposed by Vasyliunas (1983) and shown in Figure 3. In the inner magnetosphere plasma circulates around the planet in closed drift paths. Outside of some radius the drift paths are not closed. **The flux tube length stretches and the magnetic field lines pinch off to form magnetic bubbles**.

The bubbles move down tail and the short part of the flux tube snaps back to join the nearly corotating flow. This model is similar to Dungey's (1961) model for the Earth's magnetosphere converted to the jovian situation where it is powered by an internal source and not by the solar wind.

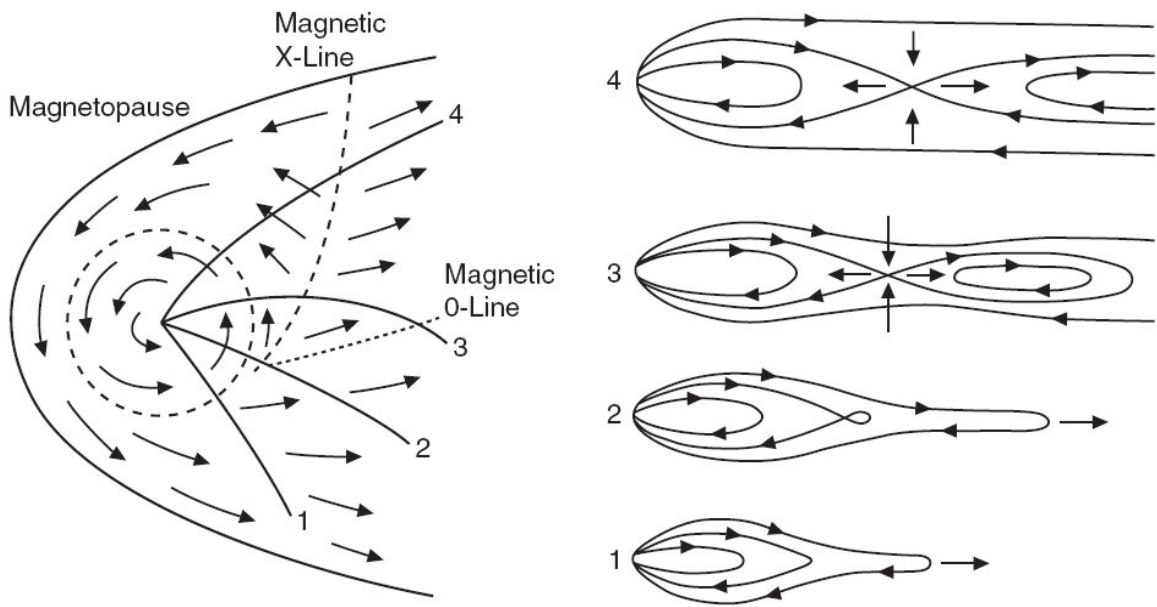

Figure 3: Jovian magnetospheric circulation model of Vasyliunas (1983). In steady state part of the circulating plasma stretches tailward and reconnects forming an island that is ejected down the tail.

The Dungey model was proposed as a steady state model to explain the time-stationary circulation of the plasma. Similarly, the Vasyliunas model is a time-stationary model. However, in both magnetospheres very interesting behavior occurs as a result of temporal changes in this circulation pattern. Thus, we will spend some time examining the how the circulation is powered and it becomes a time-varying system even through it may be uniformly driven.

Figure 4 shows isodensity contours of the Io torus derived by Bagenal (1994). The top of the figure shows the integrated density roughly along magnetic shells and over $2\pi$ radians. If one ton per second is being added to this plasma torus, a similar amount must be lost in steady state to maintain a constant density. It is difficult to lose plasma along magnetic field lines due to the centrifugal force confining the plasma to the equatorial regions and because the wave levels are too low (in this particular case, with Io as internal source) to scatter the ions out of this potential well (Russell et all., 2001). To maintain steady state the plasma must move outward at velocities given along the top of the figure. At 9 m/s it takes the plasma 3 months to move one jovian radius, but out further it requires only one month (at 31 m/s) or two weeks (at 68 m/s). Even in the region from 8 to 9 $R_J$, the plasma circulates Jupiter about 30 times as it moves radially one jovian radius.

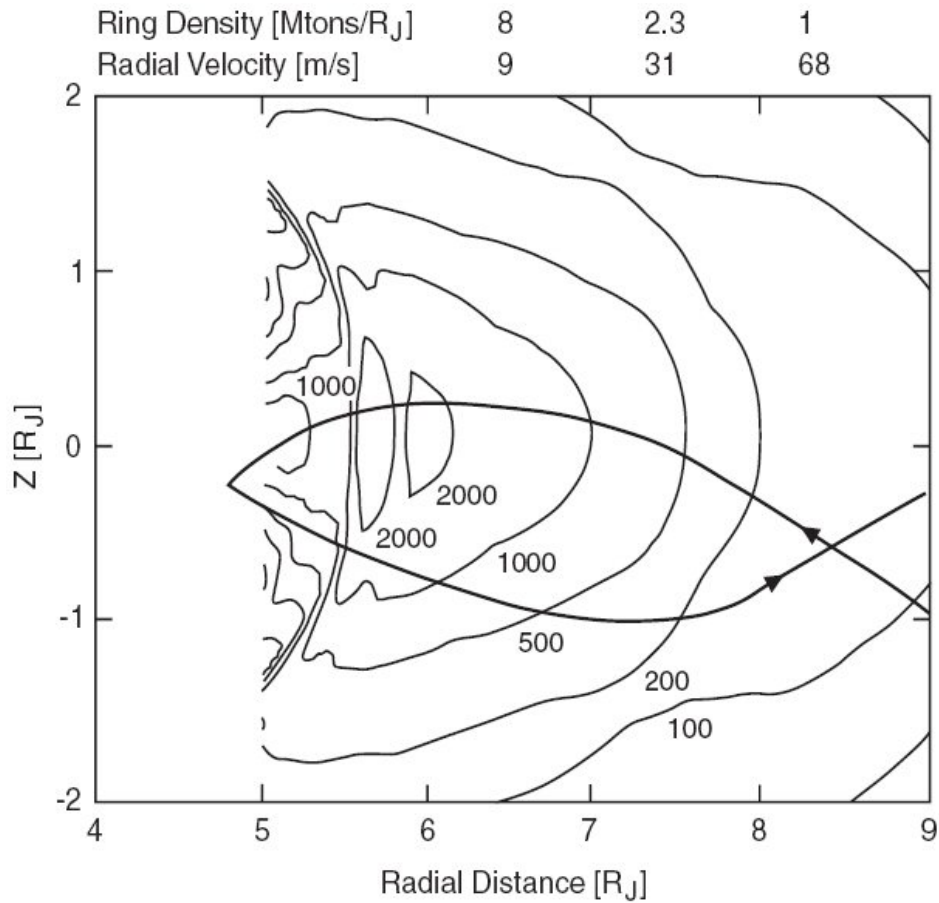
Figure 4: Isodensity contours of the Io torus (Bagenal, 1994).

There are many ways to estimate this outflow velocity. We can use observations of the Europa plume (Intriligator and Miller, 1982), the conservation of mass and stress balance in the magnetodisk region and the magnetic field normal to the current sheet. These estimates are combined with the above estimates and one made from Voyager energetic particle data to observe that the plasma is clearly driven outward by centrifugal force at a rate that increases rapidly with increasing distance (Russell, 2004). At 40 $R_J$ where the typical outflow speed is 40 km/s, a flux tube moves 10 $R_J$ in one half of a rotation.

These numbers are not unlike those implied by the Vasyliunas model shown in Figure 3. The evidence is clear that mass loading can drive a radial circulation pattern but is not immediately obvious how a steady state is maintained. The magnetic flux through Jupiter's surface is fixed, determined by the strength of the magnetic dynamo in the interior of Jupiter. However, above the surface of Jupiter this magnetic flux appears to be carried outward by the heavy ions that must be lost from the system to maintain the steady state. Again the answer lies in the reconnection process, much as occurs in the terrestrial magnetotail but with a perhaps unexpected twist.

At 60 $R_J$ behind Jupiter, at about 3 LT over a one hour period on June 17, 1997 (Russell et al., 1998), the Galileo spacecraft measured a rapid dipolarization and an increase in strength of a factor of 3 in the magnetic field. Over a 15-minute period it gradually returns to its stretched-out state. The rapid dipolarization clearly was associated with a transient reconnection event of enormous rapidity and strength perhaps propelled by the low density above and below the night time current sheet that combine with a strong magnetic field to produce a very high Alfvén velocity. The strange twist arises when reconnection moves the plasma radially inward and angular momentum conservation can make the plasma speed up in its corotational motion. Reconnection allows the magnetosphere to get rid itself of the added ions while keeping the total magnetic flux constant as it produces islands of magnetized ions that have no net magnetic flux but that can be ejected down the tail. Not coincidentally the amount of magnetic flux involved in reconnection closely matches the magnetic flux involved in the mass loading process (Russell, 2004). Russell (2004) gives also the mechanism through which the depleted magnetic flux tubes find their way back to the radius of Io where can be mass loaded again, but this phenomenon is not relevant for our present study.

### 3.4. Reconnection in Cometary Tails

The plasma tail of a comet is unique among all magnetotails in that the large scale morphology and any global changes are easily observable from the ground. As said before, magnetotails are the result of interactions between the solar wind and the magnetospheres of planets or other bodies like comets. In the case of a planetary magnetotail, the local plasma properties are observable only by in situ measurements from spacecraft. For the Earth, we can infer global changes in the magnetotail by using more than one spacecraft simultaneously. But for other planets it is difficult to see the global evolution of the magnetotail in response to variations in the solar wind due to the lack of multiple spacecraft observations. Since a comet's tail topology is globally observable, understanding cometary interactions with the solar wind can give insight into the general problem of plasma interactions in the solar system.

The most spectacular of all comet plasma tail phenomena is the disconnection event (DE), during which the plasma tail is severed from the comet's head, a process similar to the previous section's Vasyliunas (1983) model in which the same disconnection (followed by a reconnection) is responsible for the tailward plasma bubbles ejections. Although it is generally believed that the solar wind plays a role in the DE, the details of the solar wind-comet interaction responsible for initiating the tail disconnection are still under debate.

In order to understand the solar wind-comet interaction, we first must determine the solar wind conditions experienced by the comet at the time of disconnection. The popular theories for initiating plasma tail disconnection can be broadly categorized into three main classes (Brandt, 1990):

(1) *Ion Production Effects*: If the ion production rate were drastically reduced, the ionosphere of the comet could shrink temporarily and allow the magnetic field lines to slip away. The change of ionization rate in the comet could be strongly influenced by a change in the solar UV radiation.

(2) *Pressure effects*: A large scale dynamic pressure increase in the solar wind could compress the comet's ionosphere and free the magnetic field lines. If this is the correct mechanism, we would expect DEs to be associated with high speed streams or high density regions (Ip, 1980).

(3) *Magnetic Reconnection*: There are two versions of magnetic reconnection model. One is that sunward magnetic reconnection occurs as the comet crosses the interplanetary magnetic field (IMF) sector boundary (Niedner and Brandt, 1978). The other possibility is that the magnetic reconnection occurs in the tail side. According to Russell et al. (1986), the tail side reconnection might be triggered by either an interplanetary corotating shock or a high speed stream (decrease of the Alfvén Mach number), but not by an IMF sector boundary. Recently, Brandt et al. (1992) found a clear association of DEs with sector boundaries from analysis of solar wind data and comet Halley 1985-1986 data.

Even though there were not in situ observations of the solar wind at the time of the 16 March DE, Yi et al. (1994) use to study these mechanisms the solar wind data of Earth orbiting spacecrafts and other spacecrafts such as PVO, ICE, or VEGA-1. These spacecrafts have measured the direction and magnitude of the IMF, as well as the velocity, density and temperature of the solar wind. From a kinematic analysis of the wide-field photographic images from 1986 March 16 through 19, Yi et al. (1994) calculated the disconnection time to be March 16.0 (±0.1). The solar wind data around comet Halley, inferred by corotation of IMP-8 data to the comet, were such that comet Halley just crossed the IMF sector boundary.

There have been many investigations of the DEs of comet Halley in an attempt to isolate the DE mechanism. Niedner and Schwingenschuh (1987) examined the 8 March 1986 event, Lundstedt and Magnusson (1987) analyzed the 12 and 14 April 1986 DEs and Brosius et al. (1987) the 20-22 March and 11-12 April 1986 DEs. The 10 January 1986 DE was studied by Niedler et al. (1991), and Yi et al. (1993a) have discussed the 13-18 April DE. Delva et al. (1991) have used VEGA-1 spacecraft data to examine correlations between the IMF and solar wind conditions and many of comet Halley's DEs.

The formation and dynamical evolution of cometary plasma tails and magnetic boundary layers is studied by the first numerical plasma-neutral gas simulations in Konz et al. (2004). It is shown that collisional interaction between the cometary neutral envelope and the solar wind plasma leads to the formation of a magnetic barrier that separates the entire cometary body from the solar wind. The overall dynamics, in particular, of the magnetotail are governed by magnetic reconnection. In their simulation Konz et al. (2004) observed multiple reconnection events initiated by the localized excitation of anomalous resistivity. If the comet encounters a heliospheric current sheet, strong disconnection events characterize the cometary plasma tail. But even in the case of homogeneous solar wind conditions, partial disruption of the tail is triggered by dayside reconnection.

### 3.5. Emission from Wave-Particle Interactions in the Outer Magnetosphere of a Planet

In this section we will emphasize the importance of the cyclotron resonance in the observation of amplified emissions from the solar and extra-solar planets.

The cyclotron resonance (or gyroresonance) will occur between a moving charged particle and the electric component of an electromagnetic wave (Gendrin, 1972).
In real plasma any wave phenomenon has an effective bandwidth (for a distribution of particles described by some $f(\mathbf{v})$ which, in turn, has a width), leading to a wave spectrum $F(\omega,\mathbf{k})$ that is not monochromatic. We can construct these functionals by theoretical models, from plasma instability and quasi-linear theory, a set of collisionless transport quantities: diffusion coefficients, amplification factors and gain factors.

The diffusion coefficients can discriminate between several types of processes. The diffusion in configuration space can be caused by wave-particle interactions which lead to violation of one or more of the classical **adiabatic invariants** of particles trapped between turning points (mirror points) in the magnetic (geomagnetic) field.

Diffusion in configuration space occurs whenever electromagnetic fluctuations are present in the magnetosphere with spectral distribution F(ω) having significant power at frequencies such that $\omega \sim T_d^{-1}$ (breaking the third adiabatic invariant; with $T_d$ the azimuthal drift) or $\omega \sim T_b^{-1}$ (breaking the second adiabatic invariant, namely the longitudinal action; $T_b$ is the bounce period). In the later case, there may also be a **diffusion in the velocity space** because breaking the invariant associated with bounce motion usually leads to eventual precipitation of trapped particles by lowering mirror points, and thus depleting certain regions of velocity space, in turn leading to plasma instabilities which tend to replenish these depleted regions.

Diffusion in velocity space always accompanies the breaking of the first (magnetic moment) and second (longitudinal action) adiabatic invariant, which occurs for electromagnetic perturbations spanning frequencies comparable to the gyrofrequency or the bounce frequency. It readily occurs for the doppler-shifted resonance involving particles in the tail of a distribution, and leads to **pitch-angle scattering**. In the magnetosphere it is very likely that trapped radiation belt particles experience wave field perturbations over a sufficiently broad frequency spectrum that radial diffusion in configuration space occurs simultaneously with pitch-angle and energy diffusion in velocity space (Fredricks, 1975). This eventually has been discussed by Haerendel (1970), who points out that the entire source-sink properties of radial (or *L*-shell) diffusion as observed in the magnetosphere cannot be understood without assuming both configuration and velocity space diffusion simultaneously.

A second important effect due to cyclotron resonant wave-particle interactions is that of natural amplification of ULF, ELF and VLF electromagnetic waves by particle distributions having **velocity space anisotropy**. Application of this theory to ULF is found in the review of Gendrin (1972) and, to explain multiple-hop whistlers, in Liemohn (1967) and Cornwall (1965, 1966).

### 3.5.1. Whistler mode waves and turbulence

The first adiabatic invariant for trapped electrons may be destroyed by the presence of significant levels of VLF or ELF whistler mode turbulence. This was first pointed out by Dungey (1963) and Cornwall (1964). In an already classical paper, Kennel and Petschek (1966) carried out an extensive investigation in which they invoked the property of pitch-angle instability in the trapped electron population as an amplification mechanism for whistler mode waves. The model they constructed contained basically a feedback mechanism which balanced pitch-angle diffusion due to wave turbulence against precipitation and wave growth and convection out of the amplifying volume, assumed to reside near the geomagnetic equator. Thus, they were led to expressions for stably trapped limits on trapped electron flux populations. These stable limits resulted from an exact balance between loss of particles due to precipitation and wave gain due to residual velocity space anisotropy.

Using an almost identical analysis, Liemohn (1967) has shown that VLF and ULF whistlers may be amplified on field lines in near-equatorial regions at high altitudes, where pitch-angle anisotropies in particle distributions may be present. He presents this as a mechanism to largely offset ionospheric reflections, and to provide an effective guidance mechanism along field lines, since the amplification factor sharply peaks for waves with $k_\perp \to 0$. This, Liemohn points out, is an alternative to ionization ducting as a guiding mechanism for lightning whistlers, and also allows one to understand the many (up to 40) hop whistlers observed at times at ground stations.

A much-improved theory of the wave-particle interaction between magnetospheric ELF whistler mode wave turbulence and the trapped electron distributions has been published by Etcheto (1973). As first suggested by Kennel and Petschek (1966), the equilibrium trapped flux profiles and precipitation patterns (if magnetospheric dynamics ever admits the $\partial/\partial t = 0$ equilibrium) must result as a balancing of the wave turbulence spectrum, the detailed shape of trapped particle velocity distribution function, and the source and the loss mechanism for the particles and the waves.

Gendrin (1972) later pointed out that some of the concepts in the Kennel and Petschek (1966) theory of turbulent diffusion and trapped flux equilibrium were not sharply defined, for example "critical anisotropy" and "critical energy" of resonantly interacting particles. He suggested, in that paper, that one should strictly solve a set of quasi-linear equations for the self-consistent wave spectral density and trapped particle distribution function (TPDF) using a pitch-angle diffusion equation to define the TPDF, coupled to a wave kinetic equation which defines the wave spectral density function. Source terms in the pitch-angle diffusion equation are empirically defined, while loss terms are either neglected or are due to $(p, L)$ scattering in the ionosphere.

Etcheto et al. (1973) extended the qualitative ideas in Gendrin's (1972) papers, and performed self-consistent analytical and numerical solutions to the coupled pitch-angle diffusion and wave kinetic equations for an assumed functional form of the source term (injected particle spectrum), and a set of boundary conditions on wave reflection at the ionosphere and at the borders of the wave-particle interaction region. They only treated parallel-propagation ELF whistler waves ($k_\perp = 0$), an admittedly questionable assumption whose validity they discussed in their paper.

One of the basic assumptions of Etcheto et al. (1973) is that in equilibrium, the injection rate of fresh electrons ($dn_2/dt$ in their notation) must be precisely offset by the observed precipitation rate on any given $L$-shell. Thus, $dn_2/dt$, the unknown injection rate, can be related to precipitation flux, a measurable quantity. Therefore, only a model of the shape factor of the TPDF as a function of pitch angle and velocity had to be assumed by Etcheto et al. They demonstrated that their equilibrium results are relatively insensitive to this shape factor, insofar as pitch-angle distribution is concerned.

The main results of this paper are as follows. A self-consistent TPDF and wave spectrum were calculated. The peak frequency of the wave spectrum inside the plasmasphere was found to be approximately one half of the frequency at which the electrons with $v_\| = (2E_0/m_e)^{1/2}$ would resonate with whistler waves, where $E_0$ is the characteristic energy of the injected electrons. The peak intensity of the self-consistent wave spectrum was found to be directly proportional to the source intensity ($dn_2/dt$ and, thus, to the precipitation flux) and to the cold plasma density.

Etcheto et al. (1973) also found that the limiting flux expression introduced by Kennel and Petschek (1966) is a zero-order approximation, which was recognized previously by Gendrin (1972b). They found that increasing injection of fresh anisotropic electrons would indeed lead to enhanced pitch-angle diffusion and a regulation of the flux of trapped particles. However, they pointed out that there is no unique trapped flux limit, but rather they obtained an expression which is energy dependent (exponentially) and thus have quantified the ideas of Kennel and Petschek (1966), who also had stated qualitatively that their zero-order flux limitation would be invalidated for intense injection sources.

With respect to whether wave intensity or injection source intensity should yield the criterion for definition of strong and weak pitch-angle diffusion regimes, Etcheto et al. (1973) find that one can define a source intensity above which diffusion becomes so strong that the limiting flux concept no longer applies. Thus, in this sense, they answer the question by stating that the source, not the wave, intensity defines the "boundary" between strong and weak pitch-angle diffusion. The reason for this behavior appears to be related to the fact that the mean or effective anisotropy factor of the equilibrium TPDF decreases markedly in the strong diffusion limit. Thus, even through **the wave intensity increases in proportion to d$n_2$/dt**, the effective anisotropy factor defined by the self-consistent TPDF decreases (Fredricks, 1975).

The later analytical results of Etcheto et al. (1973) are consistent with the results of numerical simulations reported by Cuperman et al. (1973) that performed a particle-in-cell computer simulation of the whistler instability under various ratios of cold-to-hot plasma. They found that, consistent with the ideas of linear stability theory, initial growth rates of the whistler instability for initial pitch-angle (or temperature $T_\perp / T_\parallel >$ critical) anisotropies greater than critical (Kennel and Petschek, 1966), were indeed **enhanced by addition of cold plasma**. However, the nonlinear effects of diffusion in velocity space caused the anisotropy to decrease continuously and thus reduced growth rates. They found that, on time scales available to the simulation, all systems tested under different initial conditions on cold-to-hot plasma ratio $n_c/n_h$ relaxed to the same final anisotropy ($T_\perp / T_\parallel \sim 1.35$, which is linearly stable), and the only apparent influence of $n_c/n_h$ was to produce a final wave energy spectrum whose peak intensity was proportional to $n_c/n_h$. This trend of the numerical results is indeed consistent with the analytical results of Etcheto et al. (1973).

One of the drawbacks of the self-consistent theory of Etcheto et al. (1973) is their assumption that the wave spectrum contains only parallel propagating whistlers. As they clearly point out, **Landau effects** and **higher order resonances** with harmonics of the gyrofrequency have been neglected, and thus higher energy components ($\geq$ 200keV) of the equilibrium TPDF are not correctly treated. However, since their theory contains an "interaction length" or equatorial arc segment which is an adjustable parameter, they argue that exclusion of oblique wave normals is justified.

A theory of trapped flux profiles which does include the Landau resonance and gyrofrequency harmonic resonances for electrons interacting with oblique whistler waves has been presented by Lyons et al. (1972). The theory is not self-consistent in the sense of Etcheto et al. (1973), by not using coupled pitch-angle diffusion and wave kinetic equations in a full quasi-linear approach to find a wave spectrum consistent with their TPDF, nor did they explicitly introduce a source function for injected particles into the diffusion equation.
Lyons et al. (1972) assumed a constant level and spectral shape of ELF hiss throughout (independent of $L$) the plasmasphere. They also assumed a distribution of oblique waves with normals distributed according to a law exp(-$\tan^2 \theta / \tan^2 \theta_\omega$), where $\theta_\omega \sim 80°$ (measured away from $\mathbf{B}_0$). This distribution function of wave normals is quite flat out to about 70°, then decays rather rapidly. They demonstrate that exact distributions of wave normals are probably not important as long as wave energy is spread over a wide range, up to approximately 80°, of normal directions.

Perhaps the most recent and convincing application of Kennel-Petschek concept, combined with Haerendel's (1970) suggestions about coupling radial and pitch-angle diffusion, has been made by Lyons and Thorne (1973). These authors combined radial diffusion (source for pitch-angle diffusion) with pitch-angle diffusion (sink for radial diffusion) and Coulomb scattering (a common sink), but treated the wave spectral density function self-inconsistently, as a known function. Following the same procedures as Lyons et al. (1972), they included a broad angular distribution of oblique ELF whistler turbulence, computed diffusion coefficients for combined Landau and cyclotron harmonic resonances, averaged over bounce periods, and thus obtained electron flux distributions at several typical energies.

A main result of their treatment, even though it does not deal with the wave spectrum self-consistently, is that **the wave intensity must be significantly controlled by the convection process** (which drives radial diffusion; Fredricks, 1975). Enhanced convection apparently would be associated with enhanced spectral density of wave turbulence. Qualitatively, this is evident, since enhanced convection may reasonably be assumed to accompany a sharpening of the pitch-angle distribution of the injection spectrum, as well as an enhanced density rate $dn/dt$ of the injected particles. **This would lead to increased wave levels** and perhaps transient strong diffusion, with quick precipitation loss, and rapid restoration of more reasonable levels of trapped fluxes (Fredricks, 1975).

According to Lyons and Thorne (1973), their equilibrium flux curves are quite sensitive to fluctuations in the ratio $<E>/<B_w>$, where $<E>$ is the convection average fluctuation field during the radial diffusion, and $<B_w>$ the average whistler mode turbulence amplitude. They point out that changes in this ratio of a factor as small as 3 would be detectable over the solar cycle, as nonpersistence of the electron slot. Based on the measurements, this appears not to be the case, leading Lyons and Thorne to postulate that radial diffusion due to $<E>$ exerts effective control over $<B_w>$. During major electric storms, they do not expect inner zone flux profiles to satisfy their model, but rather to become transiently enhanced, with subsequent decay to prestorm levels.

In conclusion to this discussion of whistler turbulence, it seems fair to say that extensions of the pioneering ideas of Kennel and Petschek (1966) have now produced a large degree of qualitative understanding of the equilibrium structure of the electron belts as a result of the interaction of whistler turbulence with trapped particles. We are clearly on the verge of having a reasonably acceptable quantitative and self-consistent theory of trapped flux distributions as the ideas of Etcheto et al. (1973) are combined with those of Lyons et al. (1972).

### 3.5.2. Chorus

VLF and ELF discrete emissions from the outer magnetosphere have been observed at ground level at latitudes corresponding to magnetospheric regions just beyond the plasmapause (Carpenter, 1963). Such emissions (called chorus) were observed also by the Earth orbiting satellite OGO-5 and studied by Burton and Holzer (1974). For many years it has been postulated that such emissions are whistler-mode waves of finite duration which are triggered either by lightning generated whistlers, by powerful VLF transmissions from the ground, or are spontaneously generated by some wave-particle interaction involving cyclotron resonance. There is no disagreement among the various workers in the field that chorus emissions are generated predominantly in the vicinity of the geomagnetic equator.

Chorus emissions are important since the wave-particle interactions associated with them cause precipitation of electrons whose energy is appropriate to resonate with the VLF or ELF waves. They are furthermore interesting because they are generated by wave-particle interactions. In the following we shall discuss some of the recent theoretical and experimental work in this fertile area of research.

Discrete chorus emissions were indirectly associated with bursts of electron precipitation measured by VLF and electron detectors aboard the INJUN-3 satellite (Oliven and Gurnett, 1968) at low latitudes (~ 1000 km). Enhanced fluxes of electrons of energies bigger than 40 keV were observed along the geomagnetic field during VLF chorus bursts.

Another indirect measurement by Rosenberg et al. (1971) correlated short bursts of > 30 keV X-rays seen by detectors aboard high-altitude balloons above Simple Station, Antarctica ($L = 4.1$) and radio bursts with $\nu \sim 2.5$ kHz recorded by the ground antenna at Simple Station. Presumably, the X-rays are bremstrahlung from precipitation of electrons with primary energy ~ 60 keV and distributed over the range 30 to 100 keV. The radio bursts were observed to precede the X-ray bursts in time by some tens of seconds (~0.3 to 0.4 s). By assuming that the radio emissions are stimulated in the vicinity of the equator, and a model of plasma distribution along the relevant flux tubes passing through that region and Simple Station, the delay times between the arrival of chorus bursts and the electrons causing bremstrahlung were explained by Rosenberg et al. (1971).

Another low altitude correlation of precipitating electrons and ELF chorus using search coil magnetometer (10-1000 Hz) data and simultaneous $E > 45$ keV electron spectrometer data from OGO-6 satellite (400 to 1100 km altitude range) has been presented by Holzer et al. (1974). In this study, no attempt was made to reconcile the individual delay times between ELF chorus occurrence and precipitation. The argument used by Holzer et al. (1974) is that precipitating electrons must necessarily follow field lines from the near equatorial region in which wave-particle interactions caused by their pitch angles to diffuse into the loss cone, while the waves from that interaction region need not be guided down the same flux tube. In fact, unless ionization ducting occurs, the waves upon propagation away from the region of their unstable generation will develop a $k_\perp$, and hence an $E_\parallel$ component along $\mathbf{B}_0$. Thus, Landau damping of these oblique whistlers can occur much as in the process of oblique ion cyclotron wave damping invoked by Cornwall et al. (1971).

Even if ducting is assumed, Holzer et al. (1974) point out that the duct will effectively terminate in the high ionosphere, above their satellite, so that waves may exit the duct at angles up to their final internal reflection angle, **producing effectively an endfire waveguide antenna** irradiating the region (but not only) below the duct exit (Fredricks, 1975). Thus, electron precipitation and low altitude wave patterns need not match in general, but rather have a systematic tendency which locates wave patterns equatorward of the flux tube on which they were generated (equatorward of the precipitation pattern).

The study of Burton and Holzer (1974) presents clear evidence that chorus originates in the outer magnetosphere (beyond the plasmapause) and near the geomagnetic equator. They show that, by measuring local wave normal directions of the chorus, propagation may be either ducted or unducted. They further show that dayside and nightside chorus have distinctly different characteristics. The most important differences are:
- ➢ dayside chorus is always comprised of rising tones (risers);
- ➢ nightside chorus can be either risers or falling tones (never mixed);
- ➢ dayside chorus is found at all geomagnetic latitudes;
- ➢ nightside chorus was detected only within about 10 degrees of the geomagnetic equator;

➢ nightside chorus was only observed under magnetically active conditions and thus is most probably associated with substorm activity.

The differences in the first four characteristics indicate that the generation mechanism for dayside and nightside chorus elements may be different, and that propagation effects also may also differ. The last characteristic demonstrates that the sources of the nighside can be different from the sources of the dayside chorus. For example, the region beyond the plasmapause on the nightside consists of the plasma sheet in the "nearequatorial" region, and an extremely low density region in the lobes of the magnetotail containing polar cap field lines. The plasma sheet intensifies, thins, and moves earthward during the growth phase of substorms. Thus, one can expect the region within some 10° of the geomagnetic equator to contain an earthward convecting plasma sheet at these times. The appearance of the energetic plasma clouds at synchronous orbit (6.6 $R_E$) as seen by ATS 5 (DeForest and McIlwain, 1971) in association with substorm events indicates that copious fluxes of 10 to 20 keV electrons and protons on the order of twice this mean energy occur in this region. They have pitch-angle distributions favorable to the development of the whistler mode instability, and the energetic electrons (~10 to 20 keV) are expected to preferentially drift into the region 0000 to 0300 LT where the chorus measurements of Burton and Holzer (1974) were made. Under quiet conditions, this region of the outer magnetosphere contains little if any 10 to 20 keV electrons, so that no chorus would be expected, especially beyond the trapping boundary for > 40 keV electrons.

In any event, Burton and Holzer (1974) have shown that at least in one observed chorus generation event, the region of maximum electron pitch-angle anisotropy coincided with the origin of wave generation as determined by a wave normal analysis, and present this as experimental evidence confirming the theory of Kennel and Petschek (1966).

Structural details of chorus emissions in the outer magnetosphere have been studied extensively by Burtis and Helliwell (1969). Another, perhaps somewhat controversial, study of very fine structure of banded chorus and other emissions using electric field data has been presented by Coroniti et al. (1971). It is interesting to note that these latter authors discuss an observation from OGO-5 near 0600 LT, at a magnetic latitude near 8°, at $L \sim 6$, of a short duration sequence of falling tone chorus preceded and followed by rising tone chorus. No believable explanation of this phenomenon was given, but the observation does not necessarily refute the statement of Burton and Holzer (1974) that nightside chorus always consists, on the same pass of the satellite, of either rising or falling tones, but never both, since 0600 LT is on the boundary between night- and dayside.

### 3.6. Whistler Wave Regions in the Magnetopause Boundary Layer of the Earth

The magnetopause-magnetosheath boundary layer is characterized by tube-like structures with dimensions less than or comparable with an ion inertial length in the direction perpendicular to the ambient magnetic field. **The tubes are revealed as they constitute regions where whistler waves are generated and propagate** (Stenberg et al., 2007). It is believed that the region containing tube-like structures extend several Earth radii along the magnetopause in the boundary layer. Within the presumed wave generating regions Stenberg et al. (2007) found, using the four formation-flying Cluster spacecraft recorded data, current structures (of about or less than 20 km) moving at the whistler wave group velocity in the same direction as the waves. Furthermore, they suggest (Stenberg et al., 2005) that although the sheets of whistler waves are observed thousands of kilometers from the magnetopause, they are still directly related to the diffusion region. As an inherent property of the near-magnetosphere region, the thin layers with whistler mode waves and other small-scale structures may be the key to understand the larger scale phenomena.

The thickness of the boundary layer is several thousand kilometers being estimated from the time the spacecraft Cluster spacecraft spent in the layer and the magnetopause velocity. The magnetopause passes the different spacecraft at different times, allowing a determination of its velocity of around 100 km/s. With a transit time through the boundary layer of about two minutes we arrive at a thickness of 12000 km. In the case that the magnetopause is flapping back and forth during the two minutes, this thickness is overestimated.

Approaching the magnetopause is to be observed a change not only in the particle distributions but also in the character of the wave emissions. The structure of the waves reveals the entry into the boundary layer. Panel A in Figure 5 presents an example of such whistler mode waves detected before the Cluster spacecraft encountered the boundary layer. Although close to Nyquist frequency at 225 Hz, the fine-structure of the emission is evident. It is believed that these waves are chorus (Burtis and Helliwell, 1976) generated either close to the equatorial plane or in magnetic field minima at higher latitudes (Tsurutani and Smith, 1977). While the chorus emission is observed on closed field lines well inside the magnetosphere, the latter Whistlers are observed in the boundary layer just inside the magnetopause. The boundary layer whistlers differ from the chorus emissions in several ways. While the chorus waves appear as emissions well separated and almost regularly spaced in time, the boundary layer whistlers are more irregular. Also, chorus emissions always show time dispersion, either as risers (lowest frequencies recorded first, higher frequencies later) or fallers (the other way around). Another outstanding feature of the boundary layer whistlers is the extremely narrow-in-time but broad-banded feature extending all the way from the lowest frequencies up to the main whistler emission above 100 Hz.

Viewed as a propagation effect, the Cluster observed time dispersion is consistent with a picture where the **whistler emissions extend great distances along the magnetopause**, more than 12 $R_E$ (Stenberg et al., 2007).

The mechanism suggested by Stenberg et al. (2005) for the generation of whistler waves in the magnetopause-magnetosheath boundary layer is an electron anisotropy resulting from the intermittent and/or patchy (Stenberg et al., 2007) magnetic reconnection of the magnetospheric and magnetosheath fields through the magnetopause.

## 4. Magnetometer Measurements of Radio Atmospheric Emissions at Venus and Mars

In contrast to Earth's atmosphere Venus' atmosphere is quite dry, containing very little water. Nevertheless Venus is quite cloudy, with sulfuric acid droplets shrouding the planet some 50-60 km above the surface of the planet. Many researches have reported signals that could be produced by discharges in or from these clouds. There have been optical sightings, one from orbit of Venera 9 (Krasnopol'sky, 1983) and from the Earth (Hansell et al., 1995). There have been detections of electromagnetic waves on the landing probes of Venera 11, 12, 13 and 14 both as they descended through the atmosphere and sat on the surface (Ksanfomaliti, 1983). They were extensively studied with an electric antenna from orbit on Pioneer Venus (Taylor et al., 1979; Scarf et al., 1980). One class of ELF waves near 100 Hz had the characteristics expected for whistler propagation. Another was clearly electrostatic perhaps associated with discharges to the ionosphere (Russell, 1991). Finally, at radio frequencies Galileo reported signals similar to terrestrial radio frequency emissions (Gurnett et al., 1991).

Despite all these positive evidence, lightning on Venus remained controversial because some tests that failed to find lightning associated signals. A search for scattered light in the Pioneer Venus star sensor came up empty albeit the total search period was only a few minutes long and was spent above the side of Venus with the fewest electromagnetic pulses (Borucki et al., 1991). The photometer on the Venus balloons also did not observe flashes but it was in the clouds and not above them (Sagdeev et al., 1986). Finally, the impulsive radio frequency signals observed by Cassini on its flyby of Venus were much less frequent than those on Earth produced by lightning (Gurnett et al., 2001).

Because of this controversy, and because according to the Pioneer Venus data the signals near 100 Hz were estimated to be strong enough to be detected by the fluxgate magnetometer, an operating mode was installed in the magnetometer in which it sampled at 128 Hz with only weak attenuation in the band 64-110 Hz so that any signals in the neighborhood of 100 Hz would be captured. The method used for the magnetometers was a simultaneous sampling to within about 1 $\mu$s of each other to allow them to be subtracted from each other accurately. Natural or ambient signals appear with the same amplitude and phase on the two magnetometers. Thus, when differenced, only the noise remains. It is expected that the noise that is radiated from the spacecraft, the one that is picked up on cables and the one generated in the sensors to be of different amplitude at the two sensors.

The signals seen by the Venus Express magnetometers are very similar to what we would expect to see based on the low altitude Pioneer Venus electric field measurements (Russell et al., 2008): electric discharges in the atmosphere produce electromagnetic radiation near 100 Hz. The waveforms are bursty with individual bursts lasting from about 0.25 to 0.5 seconds. Amplitudes are generally less than 0.5 nT peak to peak. While the bursts spacing and duration are similar to terrestrial lightning, the apparent frequencies are much less. The "strokes" in the terrestrial lightning could be tens of microseconds long and producing higher frequency VLF signals in the multi-kilo Hertz range. Perhaps the difference is due to the longer path length of strokes in the Venus atmosphere that have to reach the ionosphere to discharge.

On Mars, the existence of electrical discharges generated by dust storms is an intriguing possibility. Terrestrial dust devils are known to be electrical active (Crozier, 1964), and volcanic eruptions can be accompanied by spectacular electric discharges (Anderson et al., 1965). Experimental (Eden and Vonnegut, 1973) and theoretical (Farrell et al., 1999a) investigations of dust grain electrification have found that both glow and filamentary discharges on Mars could be generated in swirling dust storms. Should such discharges be found to exist, the VLF and ELF electromagnetic radiation from them will be also present. While Farrell et al. (1999a) explored the possibility of dust devil electrification and found that filamentary discharges from Martian dust storms could be produced by the expected level of charge separation, Cummer and Farrell (1999) focused primarily on the radiation and propagation of the VLF and ELF electromagnetic energy created by such discharges. Because of the transient nature of the source (discharge current), the radiated energy is contained in short-timescale signals, termed, as we already know, radio atmospherics (or sferics).

Cummer and Farrell (1999) investigate the propagation of sferics in the spherical waveguide formed by the Martian ground and ionosphere using a numerical propagation model. They calculate the radiated VLF and ELF magnetic and electric field spectra and waveforms for a variety of source current orientations and for waveguide characteristics appropriate for Mars. What they find is that the Martian sferics has some significant differences from their terrestrial counterparts which are primarily due to the assumed lossier Martian ionosphere and less conducting Martian ground.

Martian sferics are expected to be similar for propagation under both daytime and nighttime Martian ionospheres, while, as we saw previously in this material, terrestrial sferics are known to change drastically from day to night. The modeled Martian sferics are similar to daytime terrestrial sferics in that they are not strongly dispersed, which is due to the relatively small gradient of the index of refraction in the assumed Martian ionosphere. The poorly conducting Martian ground allows nonzero tangential electric and normal magnetic fields to be observed on the Martian surface, while on Earth the almost perfectly conducting ground (at VLF and ELF) forces these components to be zero at ground level.

The importance of Cummer and Farrell (1999) model, easy to be verified by an in situ ELF and VLF dedicated Martian future spatial mission, is not only that predicts the existence of electrical discharges on Mars, but **because their propagation is controlled largely by the ionosphere and ground, it provides a Martian sferics based remotely technique for sensing these two media**. In particular, it can prove very useful in probing the large-scale Martian ground conductivity through the VLF and ELF attenuation rate. A conductivity significantly higher than that of the surface dust layer could indicate the presence of high-conductivity minerals or even subsurface ice and water. Although in Cummer and Farrell (1999) model there are assumed discharge currents similar to those on Earth, the results can be adapted to other source timescales and amplitudes. The propagation problem is linear, so sferic spectra and waveforms from more slowly varying sources can be constructed by a superposition of Cummer and Farrell (1999) waveforms.

Related work has been reported by Sukhorukov (1991) who calculated the Martian Schumann resonance frequencies and attenuation rates using an analytical approximation of the Martian ionosphere.

Sferics can be channeled through the ionosphere and observed from the magnetosphere of a planet by four competing mechanisms. The first one is by artificial ionization ducting (as in the High Frequency Active Auroral Reasearch Program – or HAARP, by ionospheric heating resulting into the launch of ELF and VLF waves up to 20,000 km; *http://www.haarp.alaska.edu*), or by natural ionization ducting, in the low atmosphere by lightning, triggering an ionization of the high atmosphere (on the same channel above the storm), manifested by sprites and jets (Su et al., 2003; Rodger, 1999). At another level we have the offset of ionospheric reflections by the amplification of whistlers in the near in near-equatorial regions at high altitudes by pitch-angle anisotropies in particle distributions (Liemohn, 1967). The third modality of sferic escape can be by transmission of stress between the ionosphere and magnetosphere by shearing a flux tube with respect to its neighbors (Strangeway et al., 2000). Finally, at least for the terrestrial ionosphere stratification case, an emission from the planet's nightside will reflect from bellow on the F region. If this radio emission is aimed toward the dayside, where the F region becomes transparent for the whistler mode, then, the next time when it will be reflected will be from above of the D or E ionospheric region, the wave propagating toward the magnetosphere.

## 5. Tries of Magnetospheric Emissions Detection for the Extra-Solar Planet Orbiting τ Boo

Among the most impressive astronomical discoveries in the past decade are the observations of Jupiter-like planets in orbit around stars similar to our Sun (Mayor and Queloz, 1995; Marcy, 1998). The field of extra-solar planet (or exo-planets) research exploded dramatically since the discovery of the first such systems in 1995. Detection methods for extra-solar planets can be broadly classified into those based on: (i) dynamical effects (radial velocity, astrometry, or timing in the case of the pulsar planets); (ii) microlensing (astrometric or photometric); (iii) photometric signals (transits and reflected light); (iv) direct imaging from ground or space in the optical or infrared; (v) miscellaneous effects (such as magnetic superflares or **radio emissions**). These detection methods have been summarized in Figure 5. Each of them have their strengths, and advances in each field will bring specific and often complementary discovery and diagnostic capabilities. **Detections are a pre-requisite for the subsequent steps of detailed physical-chemical characterization demanded by the emerging discipline of exo-planetology**.

As of December 2004, 135 extra-solar planets have been discovered from their radial velocity signatures, comprising 119 systems of which 12 are double and 2 are triple. Beyond 2015, very ambitious space (Darwin/TPF) and ground (OWL) experiments are targeting direct detection of nearby (< 100 pc) Earth-mass planets and the measurement of their spectral characteristics. Beyond these, "Life Finder" (aiming to produce confirmatory evidence of the presence of life) and "Earth Imager" (some massive interferometric array providing resolved images of a distant Earth) appear as distant visions.

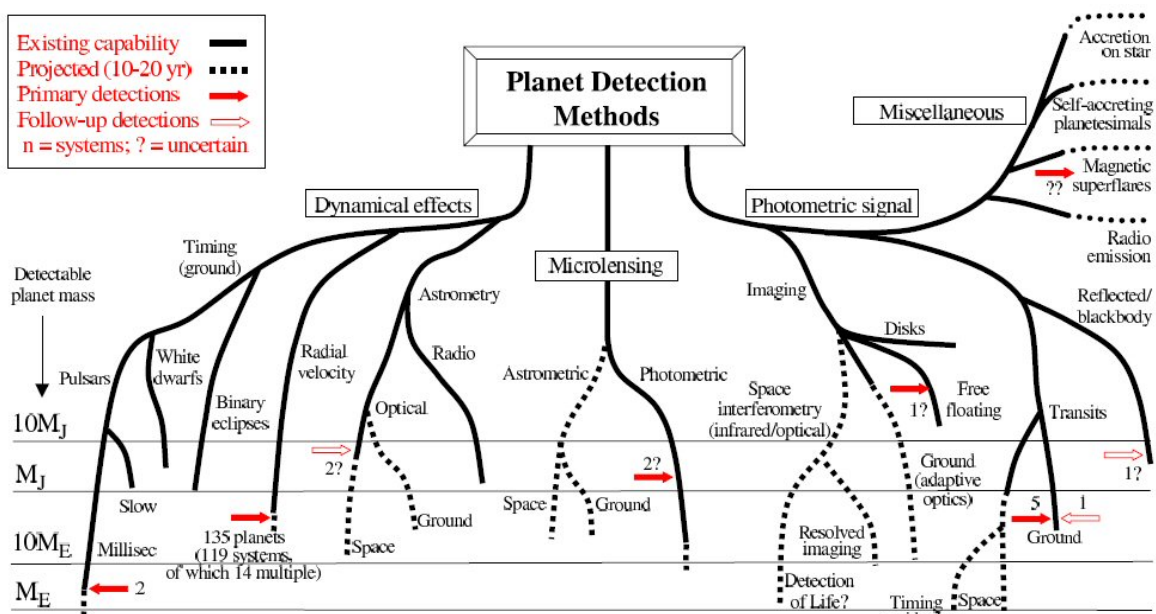

Figure 5: Detection methods for extra-solar planets (Perryman, 2000). The lower extent of the lines indicates, roughly, the detectable masses that are in principle within reach of present measurements (solid lines), and those that mght be expected within the next 10-20 years (dashed). The (logarithmic) mass scale is shown at left. The miscellaneous signatures to the upper right are less quantified in mass terms. Solid arrows indicate detections according to approximate mass, while open arrows indicate further measurements of previously-detected systems. "?" indicates uncertain or unconfirmed detections. The figure takes no account of the numbers of planets that may be detected by each method. More information and ongoing projects are given in the web page:
*http://www.obspm.fr/encycl/searches.html*

Radial velocity measurements accuracies are close to the values of around 1-3 m/s at which atmospheric circulation and oscillations limit measurement precision, implying mass detection limits only down to 0.01-0.1 $M_J$ (depending on orbital period). Detection of an Earth in the habitable zone (the range of distances from a star where liquid water can exist on the planet's surface; for a 1 $M_{Sol}$, the inner habitability boundary is at about 0.7 AU and the outer boundary at around 1.5 AU; the habitable zone evolves outwards with time because of the increasing luminosity of the Sun with age, resulting in a narrowed width of the continuously habitable zone over some 4 Gyr of around 0.95-1.15 AU) would require accuracies of ~ 0.03-0.1 m/s. Observations from space will not improve these limits, and no high-precision radial velocity measurements from space have been proposed.

The idea of "stacking up" many radial velocity observations to average the effects of stellar oscillations is appealing, but faces several complications. The first one is that even if p-mode oscillations effects can be minimized, beating amongst these modes may induce large radial velocity variations (up to 10 m/s peak-to-peak) over timescales of a few hours. The star will therefore need to be observed over several hours for each epoch (radial velocity point), a very expensive solution in terms of telescope time. The second complication that arises at the level of the long-term precision targeted is the necessity of a wavelength calibration precision from night-to-night. Reference calibration to 0.1 m/s will require further improvements in calibration techniques. With HARPS program, a precision of about 0.5 m/s is reached, as illustrated by asteroseismology results on µ Area with 250 observations each. In conclusion, a very high radial velocity precision seems possible, but at a very high cost.

There is a significant difference in the case of transiting candidates: now the period and phase are known, and with $e \sim 0$ for short period planets, a series of accumulated measurements can be used to constrain the radial velocity semi-amplitude. With HARPS at a precision of 1 m/s, for short-period planets, it is expected that limits of a few Earth masses, for $P < 10$ days, can be reached. If the transiting object is larger, then the radial velocity effect will be larger and easier to detect.

Photometric (transit) limits below the Earth's atmosphere are typically a little below the 1% photometric precision, limited by the variations in extinction, scintillation and background noise (depending on telescope aperture size), corresponding to masses of about 1 $M_J$ for solar type stars. One main challenge is to reach differential photometric accuracies of around 1mmag over a wide field of view, in which airmass, transparency, differential refraction and seeing all vary significantly. The situation improves above the atmosphere, and a number of space experiments are planned to reach the 0.01% limits required for the detection of Earth-mass planets. HST can place much better limits on transit photometry than is possible from the ground. Simulations have been made by COROT teams in order to estimate the transit detection threshold due to stellar activity. In the case of a very active star (like the Sun), it is possible if several transits are summed. In the case of COROT, 1.6 $M_E$ is detected after 10-30 transits. Another complication is false positive detections, where statistical effects, stellar activity and background binaries can all mimic transit events, and which call for independent confirmation of detections in general.

Astrometric measurements do not yet extend below the 1 milli-arcsec of Hipparcos, implying current detectability limits typically above 1-10 $M_J$. Even with the expected advent of narrow-field ground-based astrometry at 10 micro-arcsec (e.g., PRIMA), detectors would be well short of Earth-mass planets, even within 10 pc. Above the atmosphere, astrometric accuracy limits improve significantly, but is still degraded by magnetic stellar activity, spottiness and rotation, all of which giving a substantial additional contribution to the astrometric (and photometric) variability.

Microlensing searches are not limited by current measurement accuracies for Earth-mass planets, which can produce relatively large amplitude photometric signals (a few tenths of a magnitude or larger), though small amplitude signals are more frequent. The limitations of this method are rather of statistical nature: even if all stars acting as microlenses have planets, only a small subset of them would show up in the microlensed lightcurve, depending on the projected separation and the exact geometry between relative path and planetary caustic. Space measurements help significantly by reducing the photometric confusion effects resulting from observations in very crowded regions (such as the Galactic bulge) which are favored fields to improve the statistics of detectable events.

By analogy with the sun's planets, it has been predicted that these extra-solar planets will have electrical active stellar wind driven planetary magnetospheres possible capable of emitting long wavelength radio emissions (Burke, 1992; Farrell et al., 1999b; Bastian et al., 2000; Zarka et al., 2001).

The magnetic polar regions of the Earth (Gurnett, 1974) and gas giants (Burke and Franklin, 1955) are well known sites of intense **aurora related radio emissions** (Kaiser, 1989), this generated via coherent cyclotron radiation from enhanced counterstreaming electron currents (Wu and Lee, 1979). The auroral electron cyclotron radio power from Earth is approximately $10^8$ W, while for Jupiter can exceed $10^{10}$ W. Variability in this power levels is quite large (over a factor of 1000), and is exponentially correlated with the solar wind input velocity and power (Gallagher and D'Angelo, 1981). Simple linear approximations suggest that about five parts per billion of solar wind input power is ultimately converted to auroral radio power ($P_{rad} \sim \varepsilon\, P_{sw}$, with $\varepsilon \sim 5 \times 10^{-6}$) in a planetary magnetosphere.

The scaling of the radio power from magnetic planets' auroral regions has a geometric dependency on incident solar wind pressure and planetary magnetic field strength. In order to make a prediction of the radio power from Uranus prior to the Voyager 2 encounter, Desch and Kaiser (1984) correlated the auroral emissions from the known magnetic planets about the Sun and derived a "general" relationship for nominal radio power called a *Radiometric Bode's Law*. Later, it was recognized that the very intense non-Io jovian decametric radio component was also solar wind driven, and the law was modified by Zarka et al. (1997) to include this intense source and the new discoveries at Uranus and Neptune. Both laws suggest (approximately) that the auroral radiated power from planets scales directly with mass and inversely as the three halves power of their distance from the Sun.

In the mid 1990's, reports were made on the discovery of extra-solar planets about Sun-like stars in the near vicinity of our solar system. By 1998, over 9 new planets about other Sun-like stars had been identified, including an estimate of planetary mass and star-planet distance for each case (Marcy, 1998). Given these parameters and the radiometric Bode's law, a prediction of the cyclotron radiated power and frequency could be made for these known extra-solar planets (Farrell et al., 1999b).

It was considered that the best candidate for radio observations of exo-planets is the planet about Tau Bootes (F6IV star), who was predicted to have a maximum power level just above 0.1 Jansky (1 Jansky = 1 Jy = $10^{-26}$ W/m$^2$ Hz$^{-1}$), radiated cyclotron radiation between 7-70 MHz (Farrell et al., 1999b), an overestimated frequency by three to six orders of magnitude as we saw in the previous chapters.

Indirect evidence for extra-solar planetary magnetic fields comes in the form of modulation in the Ca II H and K lines of the stars HD 179949 and ν And, modulations that are in phase with the orbital periods of their planets with the smallest semi-major axes (Shkolnik et al., 2005). Though they monitored τ Boo, no similar modulations were seen. Shkolnik et al. (2005) suggest that the Ca II line modulations result from energy transport related to the relative velocity between planet and the stellar magnetosphere. Any such Ca II line modulations for τ Boo would then be suppressed because the star's rotation period is comparable to the planet's orbital period (Catala et al., 2007). Suggestively, though, the polarization observations of Catala et al. (2007) do suggest a complex surface magnetic field topology for τ Boo, consistent with a possible interaction with the planet's magnetic field.

In June 1999 and again in May 2002, the Very Large Array (VLA) near Socorro New Mexico made measurements the region near Tau Bootes in search for long wavelength radio emissions from its extra-solar planet. The VLA consists of 27 radio antennae, each of 25 m diameter, aligned in a Y-shaped configuration to yield an effective resolution of an equivalent 36 km antenna and the equivalent sensitivity of a 130 m dish antenna. During the operations, sensitivities of < 0.12 Jy were obtained at 74MHz, but no obvious signal was detected. Farrell et al. (2003) suggested three possible reasons for this lack of radio "hot spot" about Tau Bootes:
(a) the extra-solar planet was not radiating at the time of observations;
(b) the emission was not intense enough to be detected;
(c) **the emission was occurring in a frequency band outside the VLA's 74 MHz and 330 MHz system**.

In addressing (a), planetary radio source observations should be naturally periodic, where the planet is orientated with its magnetotail in our field direction, so occurring at the rotation period of the planet (tens of hours for the Sun-type star planets). Given the proximity of the planet to τ Boo, it is anticipated that it is tidally-locked with an approximately 80 hour rotation period. To compensate for this effect Farrell et al. (2003) randomly phased the measurement intervals over years, and the intervals were many hours in duration (nearly 6 hours in 2002). While this coverage is obviously incomplete, the randomly-phased snapshots were designed to minimize the effect of natural periodicity.

Considering (b), the lack of a clear signal can yield an upper limit to the radiative power from the extra-solar planet, assuming it is radiating near 74 MHz but simply too weak to detect. Specifically the radiated flux reaching Earth was considered to vary as:

$$\Phi = \frac{P}{4\pi D^2 \Delta \nu}, \qquad (1)$$

where $D$ is the sun-star distance, $\Delta \nu$ is the emission bandwidth (which is typically about $\nu_c/2$, where $\nu_c$ is the cyclotron frequency), and $P$ is the radiated power from the planet. For τ Boo, $D$ is 15.6 parsecs, $\Delta\nu$ assumed in tens of MHz and $\Phi < 0.12$ Jy, corresponding to a planetary cyclotron emission upper limit of $\sim 3 \times 10^{16}$ W.

This radiated power upper limit of $10^{16}$ W is nearly $10^5$ times greater than that of Jupiter's cyclotron emission, and initially appears to be such a large upper bound as to be of no real value. However, the planet at τ Boo is nearly four times the mass of Jupiter (4.14 $M_J$) and located at nearly 100 times closer to its parent star (semi-major axis of 0.047 AU; orbiting period 3.3 days). Given the radio scaling (only the considered auroral emission Bode's law) for magnetic planets it can be computed that the nominal cyclotron radiated power from the planet at τ Boo to be 10000 times that of Jupiter or near $10^{14}$ W. Given sun-like variability in the stellar wind from τ Boo, this nominal power can increase by another factor of 100 to near the detection threshold during intense stellar wind flow (~$10^{16}$ W peak).

Lazio and Farrell (2007) developed a likelihood method for multi-epoch observations of radio emissions from exo-planets and use it to constrain various radiation properties of the planet. Assuming that the planet radiate at 74 MHz, its typical luminosity must be less than $10^{16}$ W, unless its radiation is highly beamed into a solid angle $\Omega \ll 1$ sr. While within the range of luminosities predicted by various authors for this planet, this value is lower than recent estimates which attempt to take into account the stellar wind of τ Boo using the known properties of the star itself. Electron cyclotron maser emissions from solar systems planets is beamed, but with characteristic solid angles of approximately 1 sr illuminated. In addition, for our solar system planets, variations within the level of solar activity can amplify the cyclotron maser emission process, producing radio luminosities (and therefore flux densities) 1-2 orders of magnitude above the nominal level.

An improved version of formula (1) for the radio emissions from an extra-solar planet is (Farrell et al., 1999b; Lazio et al., 2004):

$$\Phi = \frac{P}{\Omega D^2 \Delta\nu}, \qquad (2)$$

where we consider also that the radiation is beamed in the $\Omega$ solid angle. **The luminosity, or radiated power and the emission bandwidth can be related to various intrinsic planetary properties (e.g., mass and rotation rate of the planet) or extrinsic planetary properties (emission mechanism and location of this emission – e.g., channeled sferics through ionosphere and magnetosphere, and/or emissions from wave-particle interactions in the outer magnetosphere and magnetopause-magnetosheath boundary layer, and/or reconnections in the magnetospheric circulation 1983's Vasyliunas model, and/or aurora related radio emissions)**. The standard practice has been to use empirical laws from the solar system to make predictions for $P$ and $\Delta\nu$.

Planetary radio emission has a characteristic wavelength $\lambda_c$ or frequency $\nu_c$. For the solar system planets, the cyclotron maser emission is fairly wideband with $\lambda_c / \Delta\lambda \sim 2$ ($\Delta\nu / \nu_c \sim 1/2$). In making predictions, it has been assumed that the extra-solar planetary radio emissions are comparably broadband, and, more importantly, that any observational searches would be carried out at a wavelength near $\lambda_c$ (frequency near $\nu_c$). From an observational standpoint, Lazio and Farrell (2007) incorporate in (2) a factor to take into account the possibility that a search may not have been conducted at an optimum wavelength. In lack of another option, they use a step spectrum function:

$$f_\nu(\nu,\nu_c) = \begin{cases} 1, & \nu < 2\nu_c \\ 0, & \nu > 2\nu_c \end{cases}. \tag{3}$$

A more realistic function for the emission spectrum **is the very first objective of our proposed spatial mission, constructed from observations on the solar system planets and extrapolated to the exo-planet spectrum**. For this we will require a relevant temporal observation time of our solar system planets (2-3 years) in order that to be able to differentiate the emission as function of planet type (terrestrial or gas giant), magnetosphere type (intrinsic or induced), as well as emission mechanism (or superposition of emission mechanisms and so, of spectrum functions).

Making the assumption that $\Delta\nu \sim \nu_c / 2$, Lazio and Farrell (2007) obtain a better formula for the expected flux density of an extra-solar planet when observed at a frequency $\nu$:

$$\Phi = \frac{2}{\nu_c} \frac{P}{\Omega D^2} f_\nu(\nu,\nu_c). \tag{4}$$

The methodology for searching for radio emissions from an extra-solar planet is to utilize a radio interferometer to make images of the field surrounding the planet. In the absence of a source, the pixels in a thermal noise limited image from a radio interferometer have a zero-mean normal distribution with a variance of $\sigma^2$, so that the probability density function (PDF) of obtaining a pixel with a noise intensity between $N$ and $N + dN$ is:

$$f_N(N) = \frac{1}{\sigma\sqrt{2\pi}} e^{-N^2/2\sigma^2}. \tag{5}$$

The assumption of a zero-mean distribution depends upon the image having been made without a so-called "zero-spacing" flux density, that is, without a measurement of the visibility function at a spatial frequency of zero wavelengths.

We adopt a signal model of $I = \Phi + N$ for the intensity at the location of the planet, where $N$ is the (thermal) noise in the image and $\Phi$ is the flux density contributed by the planet. We assume that $\Phi$ is constant over the duration of the observation. If this is not the case, then the PDF of $\Phi$ would have to be incorporated into this analysis. In practice, the best that one could do would be to use what is known about the radio emissions of the planets in the solar system and to develop an appropriate PDF. For the purpose of this analysis, in first approximation (and until a better choice is available), we shall treat $\Phi$ as a constant, which has the effect that we will be placing constraints on the mean level of radio emission from a planet.

A detection occurs if the pixel intensity exceeds some threshold, $I_t$. Thus:

$$p_\Phi(I > I_t \mid \Phi) = \int_{I_t}^{\infty} \frac{1}{\sigma\sqrt{2\pi}} e^{-(I-\Phi)^2/2\sigma^2} dI. \tag{6}$$

For simplicity, we define $x = I / \sigma$ and $s = \Phi / \sigma$.

Equation (4) suggests that small values of $\Omega$ would produce large flux densities. However, small values of $\Omega$ also imply that the radiation beam is unlikely to intersect our line of sight. We characterize the probability of intercept as:

$$p_\Omega(\Omega) = \frac{\Omega}{4\pi}. \quad (7)$$

The full probability of detection is then:

$$p(P, \Omega, \nu_c) = p_\Phi \, p_\Omega, \quad (8)$$

where we have made explicit the planetary parameters that we seek to determine.

Suppose that we have $n$ observations of a planet with $m$ detections, with $m = 0$ a state of detection in no epoch.

For the purpose of placing constraints on the factors in equation (4), consider first a series of trials in which the probability of detecting the planet in any single trial $p$ is the same. This case corresponds to one in which the observations are essentially identical. Then the probability of detection in the several trials is given by the binomial probability:

$$\mathcal{P}(p; m, n) = \binom{n}{m} p^m (1-p)^{n-m}. \quad (9)$$

In the case that $m = 0$, the above formula becomes:

$$\mathcal{P}(p; 0, n) = (1-p)^n \quad (10)$$

In any actual observational case, $p$ will likely vary from epoch to epoch, most likely because will have different noise levels $\sigma$. This is certainly the case if the images are obtained from different instruments. Even images from the same instrument can have different noise levels, though, depending upon the prevalence of RFI during the observations, the number of antennas used, the duration of the observation, etc.

We assume that the observations are independent, that is the probability of detecting the planet in any given observation is independent of the other observations. This assumption is certainly warranted from the observational standpoint that the noise level $\sigma$ is independent from observational epoch to epoch. Then the *joint* probability of detection is:

$$\mathcal{P} = \prod_{i=1}^{N} \mathcal{P}_i = \prod_{i=1}^{N} \binom{n_i}{m_i} p_i^{m_i} (1-p_i)^{n_i - m_i}, \quad (11)$$

for which the total number of trials $n_i$ and number of detections $m_i$ are allowed to vary from one set of trials to another. This expression is suitable for observations trials on the same exo-planet location by different experiments. For example, a three epoch observations with one trial per epoch, but with no detection, $n_i = 1$, $m_i = 0$, $N = 3$, so:

$$\mathcal{P} = \prod_{i=1}^{3} (1-p_i). \quad (12)$$

For a set of observed threshold intensities $\{x_t\}$, the likelihood function is given by equation (12):

$$\mathcal{L}(\{x_t\} \mid P, \Omega, \nu_c) = \prod_{i=1}^{3} \left[1 - p(x > x_{i,t} \mid s)\right], \tag{13}$$

where we have made explicit the parameter dependences entering into $\Phi$. In practice, the likelihood function is given in logarithmic scale.

The shape of the allowed region in a plot of the likelihood function as function of radiated power and beam solid angle (on the axes) reflects that (observing in equation (4) that the quantities $P$ and $\Omega$ are degenerate) the planet could radiate intensely but be beamed into a narrow solid angle, with a low probability of detection, or it could have a wide beaming angle but with only modest luminosity. The competing effect is that, as the beaming angle becomes smaller, the probability of it intersecting our line of sight becomes progressively smaller.

If we assume a point-source noise level $\sigma \approx 0.1$ Jy obtained in one hour for observations toward a star like $\tau$ Boo, we find that sub-dividing the observations into 10 scans (e.g., each of five minutes duration), which increases the noise level to $\sigma \approx 0.3$ Jy, leads to essentially no improvement in the constraints that can be set.

As a final remark of this section, Lazio and Farrell (2007) have great expectations from observing $\tau$ Boo planetary radio emissions by next generation long wavelength radio instruments under development: Low Frequency Array (LOFAR) and Long Wavelength Array (LWA). From our point of view these expectations are overestimated since, as we saw, the most probable planetary radio emissions are in the VLF and ELF range, much lower than the LOFAR and LWA nominal observation frequencies (tens and hundreds of MHz). The second reason for our doubt in the terrestrial arrays possibility of observing the planetary radio emissions is the implied ELF wavelength requirement (at their lower frequency end) for baselines bigger than the terrestrial diameter, but also the lack of transparency of the terrestrial atmosphere.

## 6. T.I.P.O. Space Borne Radio Interferometer
## (Scientific Experiments and Mission Requirements)

The non-transparency and severe propagation effects of the terrestrial ionosphere make it impossible for Earth based instruments to study the universe at low radio frequencies. An exploration of the low frequency radio window with the resolution and sensitivity essential to meet the scientific objectives will necessarily require a dedicated satellite based interferometer operating at these frequencies.

The 60s and 70s of the past century saw some attempts to investigate the electromagnetic spectrum below 30 MHz using space borne radio astronomy experiments onboard individual satellites (Alexander, 1971, and references therein). At these frequencies a significant improvement in the resolution of the observed phenomena is brought by the use of space borne interferometers. Basart et al. (1997) suggested a strategic plan for a space based low frequency array progressing from spectral analysis onboard a single spacecraft to a two element interferometer in Earth orbit, continuing to an interferometer array in Earth or Lunar orbit, culminating in Lunar nearside and far-side arrays. They considered it necessary to have high gain antennas and suggested the use of spherical inflatable arrays with a large number of active elements as interferometer elements for the space array. Jones et al. (2000) were the first to consider short dipoles to be suitable elements for space interferometer. They proposed a 16 element array on a distant retrograde orbit, covering 0.3-30 MHz with up to 125 kHz bandwidth of observation. These studies focused on achieving the best possible performance using the technology available then. The enormous advance in technology and the vast increase in the computing resources available have brought it within reach to plan a next generation of radio interferometers. An interferometer in this frequency range will provide a two order of magnitude improvement in both sensitivity and resolution, when compared to the existing observations from individual spacecraft.

## 6.1. Low Frequency Interferometry

Though all interferometers share the same mathematical foundations, their practical implementations change considerably with the frequency range of interest. Every few orders of magnitude in wavelength the problem of designing an interferometer changes in character and essentially evolves into a different problem with new and different aspects becoming the design drivers. For instance, the detailed design of an optical interferometer and a high frequency radio interferometer do not have much in common, even that they implement the same functional blocks. It is reasonable to expect the design considerations for a VLF interferometer to differ from those at much higher radio frequencies. Some of these considerations stem from natural causes and others from technological aspects.

The directivity, of **field of view** (FoV), of a receptor for electromagnetic waves is characterized by $\lambda/d$, where $\lambda$ is the wavelength of observation and $d$ the dimensions of the receptor. For example, the Very Large Array (VLA), one of the most successful radio interferometers, has antennas of 25 meter diameter and is most often used to observe in a wavelength range from 0.224 m to 0.0125 m (1.34 - 24 GHz). Measured in units of $\lambda$, the diameter of the antennas ranges from ~ 112 at the low frequency end to ~ 2000 at the high end, leading to FoVs smaller than a hundredth to a thousandth of a radian across. In the ELF band the $\lambda$ ranges from 10,000 km (at 30 Hz) to 100,000 km (at 3 Hz). The large wavelengths and the necessity to deploy the structure in space preclude the possibility of limiting the FoV of the receptors by using apertures many $\lambda$ in size, at least in the near future. The receptor size is expected to be smaller than the wavelength of operation for practically the entire frequency range. The FoV of individual receptors is hence expected to be very large.

Even in the absence of an intense background (as the low energies processes through which the ELF or VLF waves are emitted are not characteristic for the galactic or solar processes, in which warm or hot plasmas emit radio waves at frequencies > 1 MHz), the very large FoV imply that unlike at high frequencies where the $T_{sys}$, the equivalent noise temperature corresponding to the sum of all the contributions to the signal received at the output of an interferometer element, is dominated by $T_{rec}$, the noise contribution of the receiver electronics, at VLF frequencies, $T_{sys}$ will necessarily be very large.

For a **three dimensional sampling** the aim of interferometric imaging is to arrive at the *brightness distribution* in the sky at an observing frequency $\nu$, $I_\nu(l, m)$, from the measured visibilities, $V_\nu(u, v, w)$. For a phase tracking interferometer with a small fractional bandwidth ($\Delta\nu/\nu$), the two are related by the following expression which gives the response to spatially incoherent radiation from the far field:

$$V_\nu(u,v,w) = \int_{-\infty}^{\infty}\int_{-\infty}^{\infty} A(l,m) I_\nu(l,m) \times$$

$$\times e^{-2\pi i\{ul+vm+w(\sqrt{1-l^2-m^2}-1)\}} \frac{dl\,dm}{\sqrt{1-l^2-m^2}} \quad (14)$$

where $u$, $v$ and $w$ are the orthogonal components of the baseline. They are measured in units of $\lambda$ and forming a right handed coordinate system such that $u$ and $v$ are measured in a plane perpendicular to the direction of the phase center, $u$ pointing to the local East and $v$ to the local North. $l$ and $m$ are directions cosines measured with respect to the $u$-$v$-$w$ coordinate system and $A(l, m)$ is the antenna beam pattern. If the third term in the exponential can be ignored, then equation (14) reduces to an exact 2D Fourier transform relationship (Perley, 1999). This can usually be achieved by limiting the FoV of the antenna primary beam to a narrow enough angular region, by building a large enough aperture. This is referred to as the *small FoV approximation* and most of the existing interferometers operate in this regime. Due to the large FoVs in the ELF/VLF regime, we will be well outside this regime and hence the full 3D formalism will need to be employed for the inversion of visibility data. This complication can be avoided by choosing an interferometric array formed by **only three antennas (satellites)**, naturally situated in a plane.

For most synthesis imaging instrument in operation, due to the distribution of the antennas on a near planar surface and small FoVs, it usually suffices to decompose the baseline vectors into the $u$ and the $v$ components. For a space borne ELF/VLF interferometer with a 3D distribution of elements and very large FoVs, it will be necessary to decompose the baseline along $u$, $v$ and $w$ axes. Analogous to conventional ground based synthesis images, where the fidelity of the final image depends upon the completeness of the sampling of the $u$-$v$ plane, for an ELF/VLF space array it will depend upon the completeness with which the $u$-$v$-$w$ volume is sampled. **The constellation configuration chosen for an ELF/VLF interferometer must therefore try to achieve a good sampling of the 3D $u$-$v$-$w$ volume**, difficult but not impossible to be achieved with only three array elements.

Often, the instantaneous sampling of the $u$-$v$ plane achieved by an interferometric array falls short of the requirements for a good image of the desired part of the sky and does not provide sufficient sensitivity. Ground based instruments rely on rotation of the Earth to improve the $u$-$v$ coverage and observing for longer duration improves the sensitivity. A space borne ELF/VLF interferometer will similarly rely on the motion along its orbit and changes in baselines due to relative velocities between the constellation elements to improve its sampling of the $u$-$v$-$w$ volume and the sensitivity, **a gradual increase in the constellation baselines providing a better $u$-$v$-$w$ coverage**. Using visibilities collected over a period of time to construct a single map of the sky implicitly assumes time stationarity of the sky over the period of observation. On the time scales of operation on an ELF/VLF interferometer mission, the evolution of most astronomical sources is a non issue and we may chose to ignore the low frequency variability, weather intrinsic or introduced by propagation effects. However, the apparent position of the solar system objects, with respect to more distant objects, change rapidly. The position of the Sun, for instance, at the Earth orbit changes by about 1° per day. This implies that the sky sampled by the interferometer at different epochs corresponds to different realizations of the sky, differing in the location of the solar system objects with respect to more distant ones.

The equation (14) truly holds only for a monochromatic interferometer. As mentioned before, all practical instruments measure visibilities over a finite bandwidth, $\Delta\nu$, centered at some frequency, $\nu_c$ and, the data within $\Delta\nu$ are treated as if they were at $\nu_c$. This imprecision in handling the data leads to a gradual decrease in coherence of the signal measured at two elements with increase in $\Delta\nu$, the distance of the source from the phase center, and the baseline length. For a point source, the effect of fractional bandwidth and baseline length on the reduction in peak response ($I/I_0$, where $I_0$ is the peak response) can be conveniently parameterized in terms of a dimension parameter $\beta$ defined as $\Delta\nu/\nu_c \times \theta_c/\theta_{HPBW}$, where $\theta_c$ is the distance of the point source from the phase center measured in units of the half power beam widths for a given baseline ($\lambda_c/d$, where $\lambda_c$ is the wavelength corresponding to $\nu_c$ and $d$ is the length of the baseline) (Bridle and Schwab, 1999). Substituting for $\theta_c$, $\beta$ can be expressed as $\Delta\nu\, d\, \theta_c / c$. When $\beta = 1$ the peak response decreases to $\sim 0.8$ and further reduces to $\sim 0.5$ when $\beta = 2$. A necessary requirement for synthesis imaging is that all the measured visibilities used to reconstruct the sky brightness distribution receive coherent emission from the same physical patch of the sky.

Typically, interferometer baselines span a wide range while the bandwidths over which individual visibilities are measured is a constant. This leads to visibilities from different baselines receiving correlated emission from sky patches centered at a common spot but differing in size. At high radio frequencies this poses no problem because the FoV gets limited by the diffraction beam of the aperture before the effect of the coherence in the radiation received at different elements reduced significantly. On the other hand, at ELF/VLF frequencies, as says before, the FoV of the receptors is necessary very large. A reasonable solution to the problem of matching the patches from where correlated emission is received is **to measure visibilities over sufficiently narrow bandwidths so that even the longest baseline receives correlated flux from the entire FoV**.

The large FoV for one satellite observation means that the sky cannot be approximated as flat. In order to approach the thermal noise limit, we can use a polyhedron algorithm (Cornwell and Perley, 1992) in which the sky is approximated by many two-dimensional "facets".

At high frequencies, but also in our ELF/VLF range of interest, for most part the background emission is so weak that sky can be regarded as "cold" and the number density of radio sources (taking also into account the limited distance at which the ELF/VLF sources can be observed due to the low emission power) is such that the sky can be considered to be largely "empty". Hence, a common practice of **not imaging the entire FoV but only small parts of it from where the emission is expected is not only acceptable but also prudent**.

Mapping a large primary beam also requires caution to be exercised on another front. The geometric delay, $\tau_g$, suffered by signal arriving from different directions changes widely, from zero seconds perpendicular to baseline to $d/c$ seconds along the baseline. The process of cross correlation, however, allows for correction of only a unique geometric delay, $\tau'_g$. If the residual geometric delay, $\tau_g - \tau'_g$, exceeds $1/\nu_{ch}$ for some directions, where $1/\nu_{ch}$ is the coherence time of a band limited signal of bandwidth $\nu_{ch}$, the signal received from these directions will loose correlation. This effectively limits the FoV which can be mapped by the interferometer. Such a situation can be avoided by **correlating the signal over sufficiently narrow channel widths** so that the inequality mentioned above is never satisfied. As an example, a baseline length of 100 km requires that the channel width be $\leq 3$ kHz.

## 6.2. Propagation Effects

At the ELF/VLF frequencies, the inhomogeneous, turbulent and magnetized plasma of interstellar medium (ISM) and interplanetary medium (IPM) act like mediums with refractive index fluctuating in both space and time. The propagation of the ELF/VLF radiation from distant radio objects through this medium modifies the incident wavefronts in a considerable manner. The implications of the most relevant of these propagation effects are enumerated:

*1. Temporal broadening* – The passage of ELF/VLF radiation from compact sources through ISM and IPM leads scattering and to different travel paths to the observer. This leads to a spread in the travel time of signal, which results in smearing of transient signals. Due to the comparatively much larger distances involved, interstellar temporal broadening is much more severe than interplanetary (Woan, 2000).

2. *Absorption effects* – The free-free absorption is expected to render the ionized ISM optically thick at some turnover frequency. This frequency will be a function of both the emission measure of the medium and its electron temperature. The warm ionized medium is expected to turn optically thick for path lengths of 2000 pc at 3 MHz (Dwarakanath, 2000).

3. *Reflections, refractions and scattering close to the Sun* – Due to the large gradient in the electron density in high solar corona, a variety of unusual reflection and refraction phenomenon take place in this region. An increasingly large fraction of the solar corona becomes inaccessible to the radio waves at wavelengths below ~ 20 MHz as ray paths get reflected back into the IPM. Bracewell and Preston (1956) give an in depth discussion of this and few other interesting phenomenon. Close to the Sun, there is considerable evidence, from multi-spacecraft studies of solar burst, for the existence of anomalous beaming and large angular scattering (Lecacheux et al., 1989). We have to be aware that these propagation solar impediments affect not only the observed source radio signal, but also the data telemetry to the Earth (that, due to the lack of transparency of the Earth ionosphere in the ELF/VLF band, has to be converted to a frequency at which the atmosphere is transparent), any observation and data transmission requiring a near-Sun "avoidance region".

Because of the magnetic nature of the ISM and IPM, another effect affecting higher frequencies is the depolarization of radiation. Circular polarization, however, is not affected by Faraday rotation though the intrinsic luminosity of cyclotron processes which produce it is much lower than that of synchrotron processes (Linfield, 1996).

## 6.3. Launch and Scientific Objectives of T.I.P.O. Mission (Generalities)

In order to keep the mission economically feasible it has been chosen that the three satellites proposed for this mission (named *ELFSAT* satellites) to be deployed in space by a single rocket launch vehicle. After the launch, for which we envisaged the possibility of using the Sea Launch consortium facilities (Ocean Odyssey launch site), and the reach on orbit, the satellites packed and protected in the nose of the rocket (disposing of its own propulsion engines and gyro-stabilization) will be transferred to Lunar orbit and, finally, to Martian orbit, where the carrier will release its payload. At reach, and after their separation to 10-15 thousands kilometers, each ELFSAT satellite will serve as an interferometer element for the proposed scientific objectives. The astrophysics issues addressed by the T.I.P.O. mission are intended to study phenomena not seen at higher frequencies, namely:

*1. In situ ELF/VLF studies of Mars* and, to enumerate only some of the interesting objectives:
- scientific differences of Mars induced magnetosphere from Earth's intrinsic one;
- particle-wave located emissions and their origin (atmosphere, magnetosphere, or boundary layer);
- reconnections;
- radio spectrum function and its separation in components (with probabilities of occurrence);
- variation of radio emissions and prevalence of one creation mechanism over the other with the solar cycle;
- orbital "sounding" of the Mars subsurface layer for high-conductivity minerals, ice and water.

*2. Terrestrial observations of ELF and VLF* with emphasis on their power dependencies with distance, and having knowledge of the near-Earth missions results;
- check for the validity of the radiometric Bode's law;
- emission bandwidth;
- emission beaming;
- fine tuning of the expected flux density formula for an extra-solar planet.

*3. Observation of ELF and VLF emissions from other terrestrial type planets (Venus, Mercury)*:
- check for the dependency of the emissions with distance, planetary radius, type of magnetosphere, proximity of central star (here, the Sun), and atmospheric composition (by detection of propagating sferics);
- comparative study of Earth emissions and other planet magnetospheric emissions.

*4. Observation of giant exo-planets*, our principal candidate being the planet orbiting τ Boo, a last check of the visibility in this wavelength domain of planets other than our solar system planets, and a complementary method to the previously presented ones for determining their properties.

*5. Detection of terrestrial size exo-planets*, as a final project and the principal objective of our mission, that benefits of the results (and of a final instrumental calibration) from the previously mentioned (secondary) projects and, possibly, predictions on the star-planet distance (from a reviewed radiometric Bode's law), radius (derived from the magnetosphere size), type of magnetosphere (intrinsic or induced), transit time (from the natural periodicity of the emission), stellar activity cycles (variation of emission power for different observation epochs), and, *if* an emission housed by the planetary resonant cavity is channeled through the ionosphere into magnetosphere, further in the magnetotail and, finally in free space, and *if* this low frequency component can be analytical separated from the emissions born in the magnetosphere, we will be able to tell something about the atmospheric composition and the physical and chemical processes that take place in the atmosphere of the planet. Knowing already the spectrum, power, frequencies, etc., of the solar system planets emissions we will be able to subtract them from our observations and obtain only the exo-planetary components.

Because is expected the emission to be beamed by the magnetosphere tailward, so, relative to the satellite constellation's line of sight, observed when the planet magnetosphere shows its magnetosphere tail, in opposite direction to the stellar wind, only planets having transits will constitute possible observational targets, the same constraint applying for the observations at point 4 and drastically reduce the observation time at points 2 and 3.

For more accurate (upper limits in mJy) determination of planetary emission we require:
- astrometry accuracy to better than a beam width, setting an upper limit of 2.5σ at each epoch;
- to make use of the brightest pixel within a beam centered on the position of the star to estimate the flux density of any possible radio emission from its planet. This estimate takes into account the background level determined by the mean brightness in a region surrounding the central beam;
- to co-add ("stacked" or superposed epoch analysis) the observations (images), a technique that has been used with success to find weak sources in diverse data sets (e.g., sources contributing to the hard X-ray background, Worsley et al., 2005; intergalactic stars in galaxy clusters, Zibetti et al., 2005). Experience with images at a specific wavelength shows that the rms noise level in an image produced from the sum of $N$ images is $\sqrt{N}$ lower, as expected if the noise in the images is Gaussian random noise. Obviously, by co-adding images, we will be less sensitive to a rare, large enhancement in the planetary radio emission, but we will be more sensitive to the nominal flux density, particularly for the higher estimates for the planet's flux density (Stevens, 2005).

## *6. Discovery of new phenomena.*

Even that the choice of a Martian orbit has some realization difficulties, its scientific benefits are rather great. As problems that need to be surmounted we can mention the requirement for a bigger amount of fuel for the Earth-Mars flight than for a near-Earth orbit deployment, the second one consisting in telemetry limitations. Solutions for solving these problems are mentioned in the following sections.

Speaking of benefits, we can enumerate few of them:
- As we saw, the planetary radio emissions are "stored" and beamed in and toward the magnetosphere tail (oriented in opposite direction to the solar wind), meaning that a near-Earth mission will not see these emissions, the Earth being closer to the Sun than Mars. So, instead of limiting our objectives to Mercury, Venus and Earth, we will be able to extend our studies to observe all the inner planets, including Mars.
- In the perspective of a future Martian basis, the economic interest for determining the sub-surface composition by propagation of sferics in the Martian atmosphere is huge.
- Another benefit is that we can perform distant observations of the Earth's magnetosphere behavior under different solar wind conditions and of variations in the emitted radio power with distance for particular events registered by near-Earth missions (e.g., CLUSTER).
- The diminishing solar wind plasma density with the distance from the Sun lowers the probability of unwanted cyclotron emissions from IPM plasma interactions and of reconnections of the interplanetary magnetic field (IMF) with coronal mass ejections propagating into the heliosphere. In this manner, by placing the T.I.P.O. satellite constellation further from the Sun than the Earth, we will avoid "parasitic" signals in the receivers and fake detections of exo-planetary radio emissions.

> Last but not least, the mission will not be affected by radio frequency interference (RFI) with near-Earth artificial signals, the military interest in ELF communications being well known (e.g., Russian and US Navy submarine communications; HAARP use as space weapon).

## 6.4. Instrumentation

An honorable mission budget requires placing all three satellites in one launch, and, for this, we envisaged (taking into account that, per satellite, we computed a dry mass of approximately 270 kg) a Zenith-3SL rocket, developed for and used by the Sea Launch consortium. Another mass and fuel related deployment problem is the necessity of arriving in Mars orbit. The first natural solution to these problems is limiting the payload to minimum by equipping the satellites with the strictly required instruments in order that they accomplish their objectives. The scientific payload will consist of two instruments:

**1. Dipole Receivers**.

Each satellite will be equipped with three mutually orthogonal short dipoles, in order to record all the information in the electromagnetic field incident on the satellite. Excepting that due to their poor noise characteristics, short dipole antennas are usually not the preferred choice for receiving elements in radio astronomy, the use of three mutually orthogonal dipoles offers some advantages over the conventional use of two mutually orthogonal ones: computing all the nine ($3 \times 3$) cross-correlations per baseline allows one to construct Stokes parameters to characterize the polarization of radiation received from any arbitrary direction (Carozzi et al., 2000), as opposed to being limited to directions close to the perpendicular plane defined by the two dipoles; on being equipped with the additional ability to compute all the nine auto-correlations, the use of three orthogonal dipoles permits individual satellites to be used for direction finding of polarized sources (Ladreiter et al., 1995), a potentially useful feature for initial deployment of the constellation and for calibration; and lastly the use of independent data from a third dipole can be considered as an increase in the effective collecting area or an effective reduction in observation time needed to achieve a given sensitivity (Oberoi and Pinçon, 2008).

**2. Magnetometers**.

For studying the Martian radio emissions and, later, to subtract them from other far planetary bodies' radio emissions, we propose that each satellite to be equipped with a network of miniaturized reconfigurable ELF/VLF receivers: superconducting quantum interference device (SQUID; Farmer and Hannan, 2003). SQUIDs that measure radio frequency use Josephson junctions and are expected to sense magnetic flux in the ELF range from dc up to 1 kHz (Kawai et al., 2003). Current applications using the SQUID sensor are submarine communications, the monitoring of brain wave activity, nondestructive testing, and earthquake prediction. The major advantages of the SQUID antenna are the extremely high sensitivity, its light weight and portability. Some disadvantages of the SQUID antenna are that it has to be cryogenically cooled and that it is highly susceptible to environmental interference and noise. One solution to be applied for the reduction of environmental noise was proposed by Araya et al. (2006). For the other problem of this antenna a High-Tc SQUIDs that use the superconductor YBCO (ibco) have been developed that are cooled at 77 to 90 K by liquid nitrogen, which has a lower cost than the "basic" SQUID's liquid helium cooling and a boiling point of 4 K (which is used for the "basic" SQUID's niobium superconductor). A 77 K commercial cryocooler typically costs $40,000, is 250 × 75 × 75 mm, and weighs 4 to 8 kg. Furthermore, recent development of the Chip-Integrated Electrohydrodynamic (EHD)-Pumped Cryogenic Cooling System (by Advanced Thermal & Environmental Concepts, Inc.) could make SQUIDs extremely small and portable at 0.14 kg, 30 × 20 × 5 mm, and $500. Testing of the Micro Cryogenic Cooling in the vacuum chamber began at Goddard Space Flight Center in 2003 (Farmer and Hannan, 2003).

Our technological solution for an accurate direction and distance of the VLF/ELF triggering event is a three layered nitrogen inflating cylinder mounted on each of the three satellites, each layer containing 12 High-Tc SQUIDs disposed in three rows separated at 120° on the cylinder diameter, plus a calibration SQUID on the top of the each cylinder layer (so 39 SQUID antennas per satellite). This construction, released and inflated at the Mars orbit is not only protecting the sensitive electronics of the SQUID antennas, but also is volume and payload mass saver for the mission. Another advantage of this solution is that the inflating fluid is also cooling the antennas.

The control and the computing power for the T.I.P.O. mission will be assured by an onboard computer that enables the instruments to make effective use of the limited spacecraft resources of power and telemetry-information. Envisaging an onboard correlator, as part of the onboard computer, has as consequence that not all the spacecraft will be identical. One of them will receive the data streams from the other two and will perform onboard digital signal processing (DSP) to reduce the data volumes to be transmitted to the Earth. We refer to this satellite as the *Mother* spacecraft. In order to avoid a single point of failure in the design, it will be necessary to equip also another of the ELFSAT satellites to take up the role of the Mother.

Each of the three satellites will be provided with devices designed to neutralize the spacecraft by preventing a build up in electrical charge, similar, in principle, with the CLUSTER's Active Spacecraft Potential Control (ASPOC) device (http://sci.esa.int/science-e/www/area/index.cfm?fareaid=8). If not neutralized, the spacecraft's electrical potential can have a severe impact on the performance of the scientific instruments. ASPOC does this by emitting ions of indium into space through a small needle. These ions cancel out the electrical charge that the satellite acquires.

As components without which the satellite can not function in proper conditions and which are to be detailed if and when the here expressed ideas will be put in practice, we just have to mention the radiation and impact shielding, the engines (chemical for interplanetary travel and, after separation of the satellites at Mars orbit from a, let say, Fregat carrier, ionic), the RTG source, the combustible tanks, the cooling fluid tank, the gyro-positioning system and the star tracker. Other instruments intended to improve the telemetry, Digital Signal Processing, calibration, positioning will be discussed on short into the following.

A possible location envisaged for the mission control center is the Cheia Spatial Communication Center (Romania), acquired by RADIOCOM S.A. in 2004 and modernized in 2005.

The exo-planet detection by the T.I.P.O. interferometer will need to be confirmed by other methods (e.g., photometric or spectro-polarimetric) and for this it is intended the acquisition of a telescope to be used on the stars observed by the ELFSAT satellites and considered as positive for exo-planets.

The telemetry issues, sub-divided into those relating to intra constellation telemetry and telemetry from the Mother spacecraft to the Earth, can be simplified by doing some onboard data processing to reduce the data volumes some of the propagation effects (Oberoi and Pinçon - 2008). It is considered the option of onboard visibility computation and subsequent time and frequency averaging to reduce the data volume and increase the available bandwidth of observation. Another benefit of the telemetry to the Earth requirements is the possibility of reduction of the down-link duty cycle without compromising the observing duration, the data rate from earlier mission designs being limited by available telemetry implying that the observations could be made only for the duration of the telemetry down-link (requiring a 24 hour down-link for continuous observations). Also, in this way, the ELFSAT constellation will be able to continue its observations even when the link to the Earth will be temporary cut by the Mars or the other inner solar system bodies (including the Sun) occultation, and transmit later.

The RF bandwidth over which the visibilities are finally computed will be determined by the most limiting "bottle neck" in the data path. For this, the binary data available (as we will see in the next subsection) at individual spacecrafts will be re-sampled using 1 bit before transmission to the ELFSAT Mother satellite. The final sensitivity achieved by the design will depend on the bandwidth available for intra constellation telemetry.

The spectral width of the frequency channels is determined by the length of the longest baseline and the requirement of imaging the entire primary beam.

### 6.5. Calibration and Formation Flying

The aim of any calibration procedure is to estimate the response function of the measuring instrument. In synthesis imaging this has conventionally been done by observing astronomical sources with known properties (position, strength, structure, polarization, etc.). For this, as said before, the ELFSAT interferometer will be calibrated for the observation of extra-solar planets by the, already known at that date, solar system's planetary data.

**Prior to the launch, the dipoles will have to be calibrated in sensitivity by known ELF/VLF emitting near and far sources, in atmospheric conditions and, lately, in vacuum**.

Complex gain of the receivers will be calibrated by periodically injecting a known complex calibration signal into the signal path just after the dipoles and comparing the output from the receiver with the input signal.

Comparison with observations from the ground based instruments can be used for amplitude calibration. Also, the apparent position of few possible sources at our ELF and VLF wavelengths must be corrected by other wavelength known position. On this basis, for imaging the full data sets, low-order Zernicke polynomials can be used to model the set of position offsets and thereby infer the phase corrections to be applied.

The complex primary beam of the dipoles will have to be measured on the ground, before the launch of the mission. Mounting the dipoles on the spacecraft will considerably modify the beam shapes from those of isolated dipoles. The beam shapes must, hence, be characterized after the dipoles have been mounted on spacecraft. In flight calibration of beam shapes, if needed, will require some additional onboard functionality. It can be achieved by radiating a signal of known characteristics from one or more of the constellation members, receiving it on the others and processing it suitably. The dilution of this known signal as $1/d^2$ will indicate also the achievement of the require baselines.

The formation flying requirements relate directly to the wavelength of operation of the interferometer. Working at the longest possible wavelengths, an ELF/VLF interferometer has the least demanding formation flying requirements. While infra-red space interferometers, like DARWIN (Leger et al., 1996) require the constellation members to be positioned with relative accuracies of the order of a centimeter, for an ELF/VLF interferometer it is not necessary to fly the constellation in a rigid pre-defined configuration at all, but actually better when we take into account the coverage of the *u-v-w* volume. Departures of individual spacecrafts from their intended positions, by small fractions of the characteristic length scale of the configuration, do not degrade the performance on the interferometer in any significant manner. The mission requires the baselines to be calibrated to an accuracy of ~ 0.1 $\lambda$ at the smallest wavelength of observation. It is however not necessary to know the relative positions of the constellation members to this accuracy in real time. Real time baseline accuracies only need to be sufficient to ensure that the correlation between the signal received at different satellites is not reduced significantly. **Real time accuracies of the order of a few tens of meters to one kilometer will be quite acceptable**. The final accuracies, made available by the *post facto* orbit reconstruction, must however meet more stringent requirements. Once the baselines are known to final required accuracies, offline corrections can be applied to the computed visibilities. Being far away from the Earth, it will not be possible for the mission to make use of the Global Positioning System (GPS) satellite network for locating the satellites. **An onboard ranging and direction finding system will be required**.

The intra constellation ranging and direction finding data is insensitive to an overall rotation of the array, which will need to be determined by independent means. **There is also a need to align the relative orientations of constellation members (attitude control), to ensure that the dipoles on different spacecrafts are pointed in the same directions**. A **star tracker unit onboard every constellation member** will be used to serve both these purposes. As the FoVs of dipoles are huge, attitude control of the order of a degree is probably an overestimated task.

The time synchronization requirements for the constellation do not pose a challenge. As the design does not require a local oscillator, there is no requirement to distribute a phase locked signal to all spacecrafts. The only requirement for time synchronization comes from the coherence time of the signal, which is proportional to $1/\nu_{ch}$. Given that the relative positions of the spacecrafts will be known to a few hundred meters, it will be preferable to time tag the time series of spectra from different constellation members on the Mother spacecraft, taking into account the changing light travel time and the fixed delay due to the signal processing.